\documentclass[
superscriptaddress,
reprint,
prb,
amsmath,
nofootinbib,  
amssymb,
bibnotes
]{revtex4-2}

 \newcommand{\bcen}{\begin{center}}
 \newcommand{\ecen}{\end{center}}
 \newcommand{\btab}{\begin{tabular}}
 \newcommand{\etab}{\end{tabular}}
 \newcommand{\bdes}{\begin{description}}
 \newcommand{\edes}{\end{description}}

 \newcommand{\beq}{\begin{equation}}
 \newcommand{\eeq}{\end{equation}}
 \newcommand{\bea}{\begin{eqnarray}}
 \newcommand{\eea}{\end{eqnarray}}

 \newcommand{\half}{\frac{1}{2}}
 \newcommand{\bary}{\begin{array}}
 \newcommand{\eary}{\end{array}}
 \newcommand{\benum}{\begin{enumerate}}
 \newcommand{\eenum}{\end{enumerate}}
 \newcommand{\bitem}{\begin{itemize}}
 \newcommand{\eitem}{\end{itemize}}

 \newcommand{\bsig}{\mbox{\boldmath $ \sigma $}}

 \newcommand{\balp}{\mbox{\boldmath $ \alpha $}}

 \newcommand{\bdelta}{{\boldsymbol{\delta}}}

 \newcommand{\bOne}{{\boldsymbol{1}}}
 \newcommand{\ba} {{\boldsymbol{a}}}
 \newcommand{\bb} { \mbox{\boldmath $b$}}
 \newcommand{\bc} { \mbox{\boldmath $c$}}

 \newcommand{\bk} { {\boldsymbol{k}} }
 \newcommand{\bl} { {\boldsymbol{l}}}

 \newcommand{\bq} {{ \boldsymbol{q}} }
 \newcommand{\br} { {\boldsymbol{r}}}
 
 \newcommand{\bt} {{\boldsymbol{t}}} 
 \newcommand{\bu} { {\boldsymbol{u}} }
 \newcommand{\bv} { \mbox{\boldmath $v$}}
 \newcommand{\bw} { \mbox{\boldmath $w$}}

 \newcommand{\bz} { \mbox{\boldmath $z$}}
 
 \newcommand{\bA} { \mbox{\boldmath $A$}}

 \newcommand{\bG} { {\boldsymbol G}}

 \newcommand{\bJ} { {\boldsymbol{J}}}
 \newcommand{\bK} { {\boldsymbol{K}} }

 \newcommand{\bP} { \mbox{\boldmath $P$}}

 \newcommand{\bU} { {\boldsymbol{U}}}
 \newcommand{\bV} {\boldsymbol{V}}
 \newcommand{\bW} { \mbox{\boldmath $W$}}

 \newcommand{\bzero} { {\boldsymbol{0}}}
 
 \newcommand{\cG}{\mathscr{G}}

 \newcommand{\dou}{\partial}

 \newcommand{\llangle}{{\langle \! \langle}}
 \newcommand{\rrangle}{{\rangle \! \rangle}}
 \newcommand{\disave}[1]{{\llangle #1 \rrangle}}
 
 \newcommand{\D}[1]{\textup{d}{#1}} 
 \newcommand{\grad}{{\boldsymbol \nabla}}

 \newcommand{\mean}[1]{\langle #1 \rangle}
 
 \newcommand{\bra}[1]{{{\langle #1 |}}}
 \newcommand{\ket}[1]{{| #1 \rangle}}

 \newcommand{\eqn}[1]{eqn.~(\ref{#1})}

 \newcommand{\fig}[1]{fig.~\ref{#1}}
 
 \newcommand{\ttd}[1]{{\color[rgb]{1,0,0}{\bf #1}}}
 \newcommand{\red}[1]{{\color[rgb]{1,0,0}{\protect{#1}}}}
 \newcommand{\blue}[1]{{\color[rgb]{0,0,1}{#1}}}
 \newcommand{\green}[1]{{\color[rgb]{0.0,0.5,0.0}{#1}}}

 \newcommand{\mOne}{\mathbf{1}}

 \newcommand{\calS}{\mathcal{S}}

 \newcommand{\Integers}{{\mathbb{Z}}}

 \newcommand{\ci}{\mathbbm{i}}

 \newcommand{\VBS}[1]{{ \color{BrickRed} (\textsf{VBS}) #1}}
 
 \DeclareMathOperator{\tr}{tr}
 \DeclareMathOperator{\sgn}{sgn}
 
 \newcommand{\mbeta}{{\upbeta}}
\newcommand{\oibook}[1]{}

\renewcommand{\half}{\frac{1}{2}}

\newcommand{\eqnref}[1]{Eq.~(\ref{#1})}
\newcommand{\figref}[1]{Fig.~\ref{#1}}
\newcommand{\sfigref}[2]{Fig.~\hyperref[#1]{\ref{#1}#2}}

\newcommand{\dontshowthis}[1]{{ }}

\usepackage{marginnote}
\usepackage{slashed}
\usepackage{graphicx, color, soul}
\usepackage{float}

\usepackage[breaklinks, colorlinks, citecolor=NavyBlue,linkcolor=NavyBlue,urlcolor=NavyBlue]{hyperref}

\usepackage{cancel}
\usepackage{float}
\usepackage{physics}
\usepackage[usenames,dvipsnames, pdftex]{xcolor}
\usepackage{mathtools}
 \usepackage{mathrsfs}
 \usepackage{amssymb} 
 \usepackage{marvosym}
 \usepackage{pifont}
 \usepackage{mathbbol}
 \usepackage{wrapfig}
 \usepackage{upgreek}
 \usepackage{bbm}
 \usepackage{yfonts}
 \usepackage{xspace}
 \usepackage{lineno}
 \usepackage{tikz}
\usepackage[most]{tcolorbox}
\usepackage{pgfplots}
\usepackage[mathscr]{eucal}

\usepackage[sep=5pt, offset=1.25em]{simpler-wick}

\newcommand{\paraheading}[1]{\noindent{\em #1:}}

\newcommand{\SB}[1]{{\color{NavyBlue} [SB]#1}}

\newcommand{\conc}{{\mathfrak{c}}}
\newcommand{\IPR}{{\mathscr{P}}}
\newcommand{\barz}{{\bar{z}}}
\newcommand{\barw}{{\bar{w}}}
\newcommand{\barPsi}{{\bar{\Psi}}}
\newcommand{\barJ}{{\bar{J}}}
\newcommand{\barF}{{\bar{F}}}
\newcommand{\barK}{{\bar{K}}}
\newcommand{\barQ}{{\bar{Q}}}
\newcommand{\bara}{{\bar{a}}}
\newcommand{\smsec}[1]{{Appendix \ref{#1}}}
\newcommand{\IISc}{Centre for Condensed Matter Theory, Physics Department, Indian Institute of Science, Bengaluru 560012, India}
\newcommand{\IITB}{Department of Physics, Indian Institute of Technology Bombay, Powai, Mumbai 400 076,  India }
\begin{document}
\title{Critical States of Fermions with ${\mathbb{Z}}_2$ Flux Disorder }
\author{Hiranmay Das}\email{hiranmayd@iisc.ac.in}
\affiliation{\IISc}
\author{Naba P. Nayak}\email{nabaprakash@iitb.ac.in}
\affiliation{\IITB}
\author{Soumya Bera}\email{soumya.bera@iitb.ac.in}
\affiliation{\IITB}
\author{Vijay B.~Shenoy}\email{shenoy@iisc.ac.in}
\affiliation{\IISc}
\begin{abstract}
Motivated by many contemporary problems in condensed matter physics where matter particles experience random gauge fields, we investigate the physics of fermions on a square lattice with $\pi$-flux (that realizes Dirac fermions at low energies), subjected to flux disorder arising from a random ${\mathbb{Z}}_2$ gauge field that results from the presence of flux defects (plaquettes with zero flux). At half-filling, where the system possesses BDI symmetry, we show that a new class of critical phases is realized, with the states at zero energy showing a multifractal character. The multifractal properties depend on the concentration $\mathfrak{c}$ of the $\pi$-flux defects and spatial correlations between the flux defects. These states are characterized by the singularity spectrum, Lyapunov exponents, and transport properties. For any concentration of flux defects, we find that the multi-fractal spectrum shows termination, but {\em not freezing}. We characterize this class of critical states by uncovering a relation between the conductivity and the Lyapunov exponent, which is satisfied by the states irrespective of the concentration or the local correlations between the flux defects. We demonstrate that renormalization group methods, based on perturbing the Dirac point, fail to capture this new class of critical states. This work not only offers new challenges to theory, but is also likely to be useful in understanding a variety of problems where fermions interact with discrete gauge fields.
\end{abstract}

\maketitle

\section{Introduction}\label{sec:Intro}
Gauge theories are one of the foundational frameworks to describe nature at small length and time scales, as evidenced by their success in particle physics. Surprisingly, this same framework turns out to be of central importance in describing quantum phenomena even in condensed matter systems~\cite{Kogut1979,Lee2006}. In particular, matter coupled to gauge fields makes an appearance in the description of frustrated quantum magnets and spin liquids~\cite{Wen2002, Wen2017}. Indeed, exotic phenomena such as deconfined quantum criticality owe to the emergence of gauge fields at low energies of such systems~\cite{Senthil2023}. Another notable example of such emergent gauge fields in a solvable model of a quantum spin system was introduced by Kitaev~\cite{Kitaev2006}, motivated by the desideratum of producing quantum systems with non-Abelian anyons, which has also stimulated a great deal of experimental work~\cite{Hermanns2018,Trebst2022}. In this model (in a specific regime of parameters), the matter fields are Majorana particles that move in the background of $\Integers_2$ gauge fields on a honeycomb lattice. The $\Integers_2$ flux per plaquette is unity in the ground state, and excited states have some plaquettes with negative unity $\Integers_2$ flux. Thus, the matter particles move in the background gauge fields with a disorder in the flux pattern. It is particularly important to understand the physics of this system, as experiments probe the high-temperature physics where the flux disorder emerges naturally~\cite{Perkins2019,Udagawa2021}. Fermions coupled to gauge fields are also of interest in studying novel phases and phase transitions~\cite{Gazit1,Gazit2,Hermele2004a,Nandkishore2012,Maciejko2017,Konig2020,Parasar2023}.

The phenomenon of matter particles moving in the background of gauge fields poses several interesting fundamental questions. In a regime (or in a class of problems) where the dynamics of the gauge fields are slow compared to the matter particles, how does the disorder induced by the gauge fields affect the physics of the matter particles? Does it lead to localization? What are the signatures of these effects on the properties observed in experiments (as already noted above)? 

With these motivations, in this paper, we study the problem of fermionic particles moving in the background of disordered $\Integers_2$ gauge fields on a square lattice. The clean system that we consider on the square lattice has $\pi$-flux through each plaquette arising from an $\Integers_2$-gauge field that lives on the links and couples to fermions via their hopping amplitudes. At half filling, or vanishing chemical potential, the low-energy description of the fermions is obtained using the Dirac equation. The disorder in the gauge field arises from randomly occurring flux defects, i.~e, plaquettes with $0$ flux (further details below). 
The disorder realized by random $\pi$-flux defects of fermions on a square lattice at half-filling falls in the symmetry class BDI~\cite{Altland1997}, while away from half-filling the system belongs to the AI class. Physics of disorder in chiral classes (such as BDI) has received a great deal of attention~\cite{EversMirlin2008}, following the seminal work of Gade and Wegner~\cite{Gade1991,Gade1993} who showed using a field theoretical approach that the states at zero energy (band center, i.~e., chemical potential at half filling) remain critical owing to an extra $U(1)$ factor that prevents the flow of coupling constant of the resulting sigma model. The theory also predicts a diverging density of states at zero energy, which has received considerable attention~\cite{Gade1991,Gade1993,Ludwig1994,Motrunich2002,Mudry2003}. These developments are particularly important in uncovering the physics of graphene with vacancies~\cite{Hafner2014,Sanyal2016}, where it was shown that the wavefunctions at zero energy show freezing multifractality~\cite{Hafner2014}.  An interesting question of whether a localization transition is possible in the chiral classes was addressed in~\cite{Konig2012} and has also been recently studied~\cite{Karcher2023,Nayak2024}. These developments lead to the specific questions related to those raised in the previous paragraph: What is the nature of states that result from flux disorder? Do they exhibit freezing? How does this depend on the concentration $\conc$  and the local correlations of the flux disorder? Is there a ``transition'' from a multifractal state to a frozen state with tuning of $\conc$?

In this paper, we uncover the nature of these states at half-filling by using a combination of numerical diagonalization, transfer matrix methods, and transport calculations. We find that these states are critical with a multifractal nature, displaying a termination behavior rather than freezing. This is seen in the inverse participation ratio of states, in the transfer matrix correlation length, and in the transport. The character of the multifractal state depends on $\conc$, and in the range studied in this paper $\conc = 0.005$ to 0.5, we do not find any freezing, although the multifractal character depends on the value of $\conc$ and the nature of local correlations between the flux defects. Remarkably, we find that the class of critical states realized have a common description -- we uncover that the d.~c.~conductivity (which is order $e^2/h$) is related  to the Lyapunov exponent, and systems with different defect concentrations and short-range defect correlations all follow the same relationship. We show that perturbative renormalization group methods starting from the Dirac theory fails to uncover the physics and point out that non-perturbative coherent multi-defect scattering is essential to capture the physics observed in this system. At finite energies (where the symmetry reduces to AI), we find that the states are localized (we do not discuss non-zero energies in this paper), consistent with recent work on the honeycomb lattice~\cite{Zhuang2023}.

The next section \ref{sec:Model} introduces the model, and section~\ref{sec:NumericalResults} contains the key numerical results. This is followed by a discussion in section \ref{sec:Theory}, and the paper is concluded in section \ref{sec:Summary}. Appendices \ref{SM:sec:SingleDefect}-\ref{SM:sec:Tmatrix} contain discussion of the details of the calculations, both numerical and analytical.


\begin{figure}
\centerline{\includegraphics[width=0.45\columnwidth]{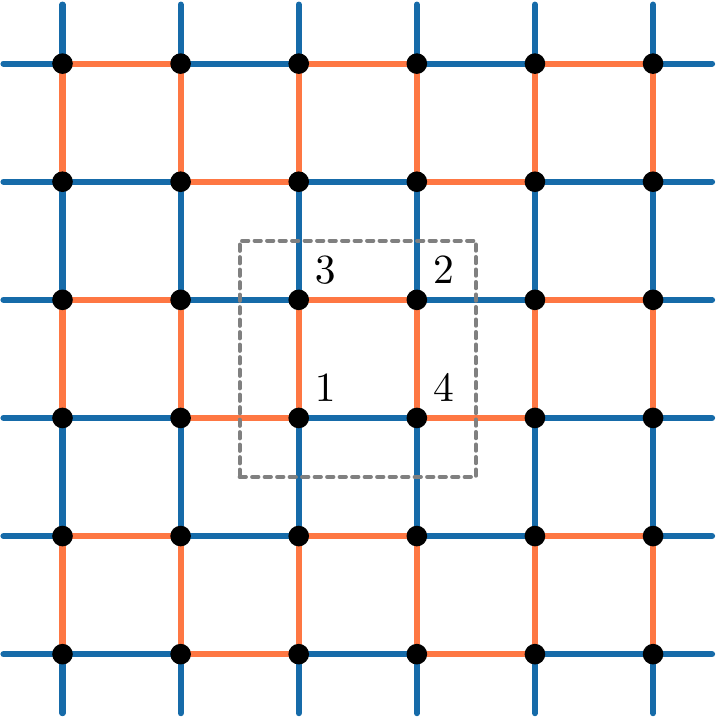}~~~~\includegraphics[width=0.45\columnwidth]{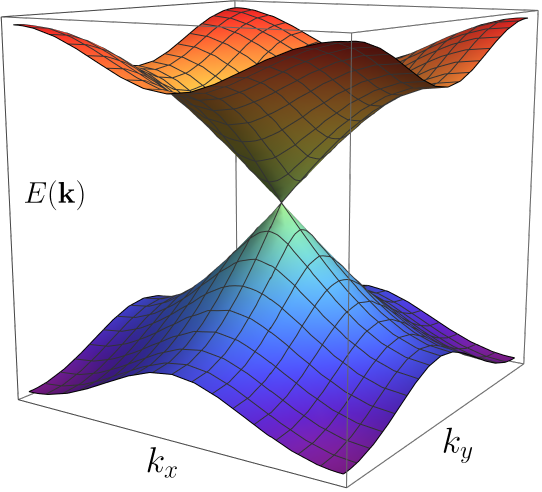}}
    \caption{{\bf Square lattice with $\pi$-flux:} Left: Unit cell consisting of four labeled sites shown as a dashed square. The orange-colored links have hopping $t=+1$ while the blue links have hopping $-1$.  Right: Band dispersion showing four bands (each two-fold degenerate) touching at a Dirac point at $\bk=0$ in the Brillouin zone.}
    \label{fig:schematic_of_model}
\end{figure}

\section{Model and Problem Statement}\label{sec:Model}


\begin{figure}
    \centerline{\includegraphics[width=0.66\linewidth]{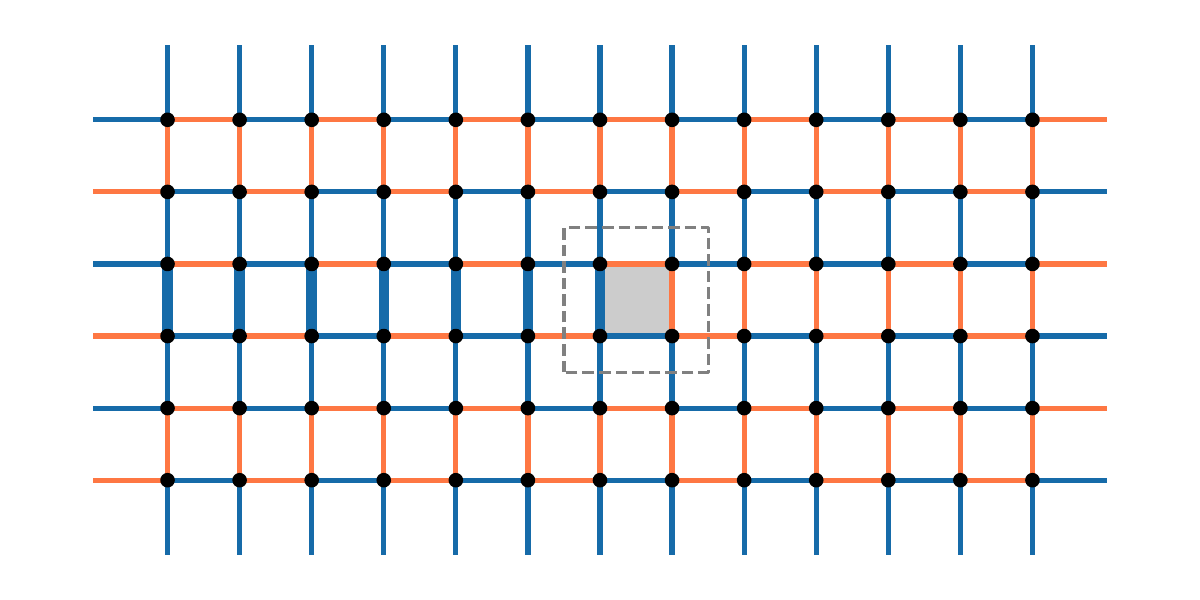}
    \includegraphics[width=0.33\linewidth]{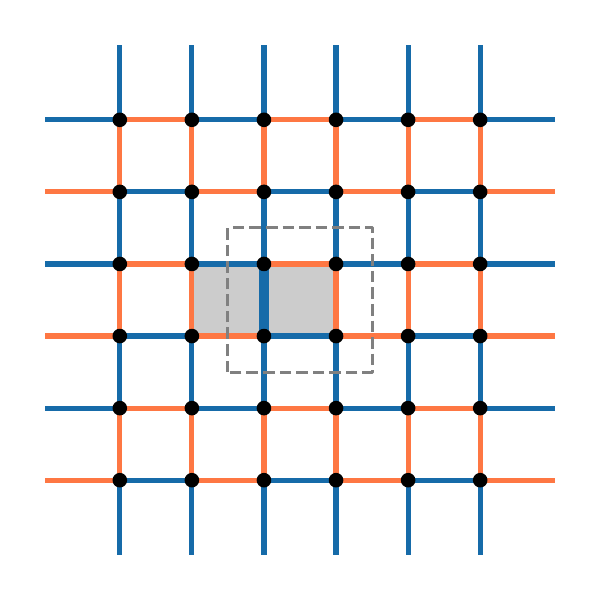}}
    \centerline{~~~~~~~~~~~~(a) Monopole ~~~~~~~~~~~~~~~~~~~~~~~ (b) Dipole}
    \centerline{
    \includegraphics[width=0.33\linewidth]{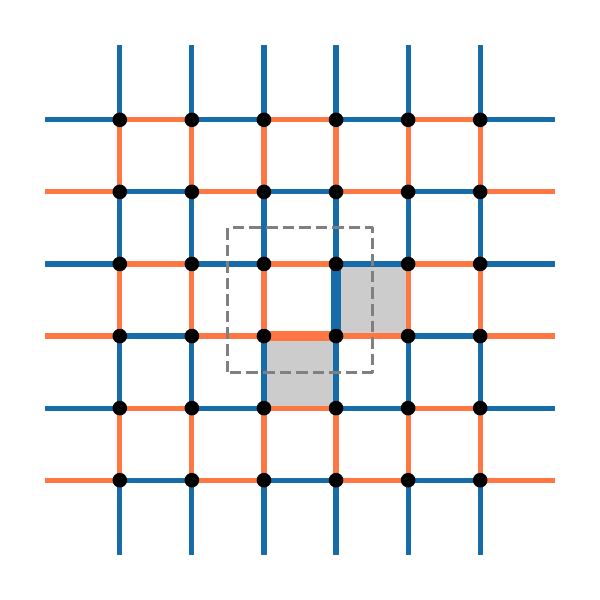}
    \includegraphics[width=0.33\linewidth]{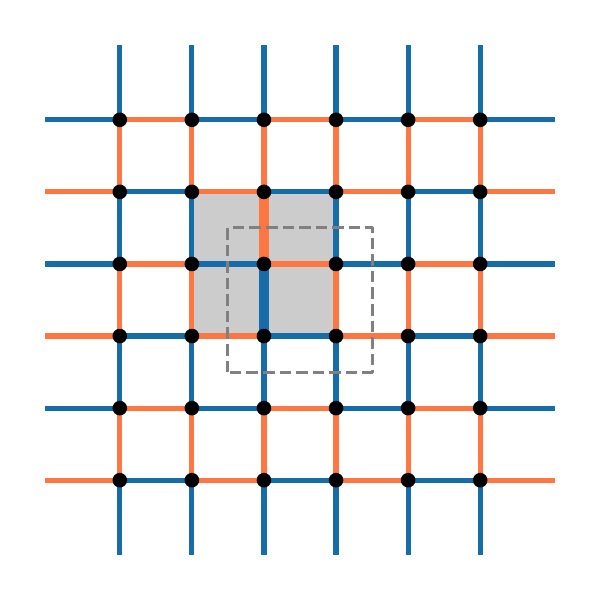}
    \includegraphics[width=0.33\linewidth]{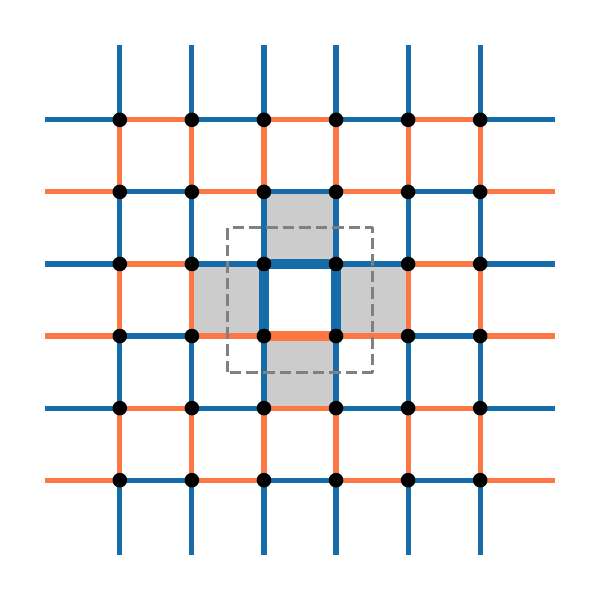}}
    \centerline{(c) Tilted dipole ~~~~~~ (d) Quadrupole ~~~~~ (e) Greek cross}
    \caption{{\bf Defect cluster configurations:} Flux defects (plaquettes with zero flux) are obtained by changing the sign of gauge fields along the links. Different panels show the different defect clusters discussed in the text, and display the pattern of the flipped gauge fields required to produce these defect clusters.}
    \label{fig:single_clusters}
\end{figure}

The model adopted in this paper is shown in \figref{fig:schematic_of_model} has a 4 site unit cell (sites labeled $1,2,3,4$ ) on a square lattice, with a unit repeat distance between unit cells.  The operator $c^\dagger_{Ia}$ creates  a fermion at the $a$ site (also dubbed as ``flavor'') in the unit cell $I$, and the Hamiltonian is given by
\begin{equation}
 \label{eqn:Hamiltonian}
H = - \sum_{I,a,b,\delta} t_{ab}(I,\delta) c^\dagger_{(I+\delta)a} c_{I b}
\end{equation}
where $\delta \in \{e_x,e_y\}$  is a nearest neighbour vector,  the hoppings $t_{ab}(I,\delta)$ are chosen to reflect the presence of an $\Integers_2$ gauge field, in that $t_{ab}(I,\delta) = z_{ab}(I,\delta) t$ where $t>0$, and $z_{ab}(I,\delta) = \pm 1$ is a static $\Integers_2$-gauge field. The reference configuration (where the gauge field is not disordered) is chosen such that every plaquette of the square lattice has $\pi$-flux through it. Our choice of $z_{ab}(I,\delta)$ that realizes this is shown in the left panel of \figref{fig:schematic_of_model}. 
The system for has time-reversal symmetry, an antilinear symmetry that acts as 
\begin{equation}\label{SM:eqn:TR}
    \Theta c^\dagger_{I a} \Theta^{-1} = c^{\dagger}_{Ia}
\end{equation}
with $\Theta^2 = 1$.
We focus zero chemical potential in the many-body setting, i.~e., the system at half-filling, and this endows the model with a sublattice antilinear symmetry
\begin{equation}
{\cal S} c^\dagger_{I a} {\cal S}^{-1}= s_a c_{Ia} 
\end{equation}
where 
\begin{equation}
s_a =  \begin{cases}
1 & a = 1, 2 \\
-1 & a = 3, 4.
\end{cases}   
\end{equation}
These symmetries place the model in the BDI symmetry class~\cite{Altland1997,HatsugaiWen1997,Guruswamy2000}.

The infrared physics of this half-filled system can be captured by a continuum field theory starting with study of the band structure of the system. There are a total of four -- two sets of two-fold degenerate bands -- that touch at zero energy at the $\Gamma$-point $\bk = \bzero$ of the Brillouin-zone ($\bk \in [-\pi,\pi]^2$), resulting in a Dirac-cone-like feature (see \figref{fig:schematic_of_model}). The low-energy field theory~\cite{HatsugaiWen1997,TanakaHu2005}
\begin{equation}\label{eqn:Dirac0}
    {\cal H}_{0} = \int \D{^2}{\br} \, \Psi^\dagger(\br) \alpha_i ( - \ci \dou_i) \Psi(\br)
\end{equation}
describing this is obtained by well known techniques, where $\Psi$ is the $4 \times 1$ column vector consisting of the long wavelength fermionic fields corresponding to the flavor labels $a$, $\br$ is the position vector, $\dou_i$ is the spatial derivative along the directions $i=1,2$, $\ci = \sqrt{-1}$ and $\alpha_i$ are the $4\times4$ Dirac matrices
$$
\alpha_1 = - \sigma_2 \otimes \sigma_1; \;\;\; \alpha_2 = \sigma_2 \otimes \sigma_3
$$
where $\sigma_{0,1,2,3}$ are the $2\times2$ Pauli matrices ($\sigma_0$ denotes the $2\times2$ identity matrix).  The Dirac matrices obey the following Clifford algebra relation
$$
\alpha_i \alpha_j + \alpha_j \alpha_i = 2 \delta_{ij}.
$$

The central aim of this study is to investigate the role of the disorder arising from the gauge fields $z_{ab}$.
Disorder in the gauge field obtains from the introduction of $\pi$-flux defects, i.e., plaquettes that have $0$ flux through them, by flipping the signs of the $z_{ab}(I,\delta)$ as illustrated in \figref{fig:single_clusters}.  We characterize the disorder by specifying two parameters. First, is the overall concentration of the flux defects which is equal to the number of flux-defects divided by the number of plaquettes in the system. The second is the short-range correlation between the positions of these flux defects -- we consider cases where the defects appear in well-correlated clusters, realizing the given overall concentration $\conc$. We consider five types of defect clusters as shown in \figref{fig:single_clusters}: (i) Monopole defects (see \figref{fig:single_clusters}(a)) (ii) Dipole defect cluster(see \figref{fig:single_clusters}(b)) where flux defects are on adjacent plaquettes (plaquettes that share a link), randomly along the $1 (x)$ or $2(y)$ directions (iii) Tilted dipole defect cluster (see \figref{fig:single_clusters}(c)) where the flux defects are the nearest plaquettes (that share only a site) at $\pm 45^\circ$ (orientation is random) (iv) Quadrupole defect cluster where four nearest plaquettes, each of which share a link with two others, host flux defects (see \figref{fig:single_clusters}(d)) (v) Greek cross defect cluster where the flux defects occupy nearest four plaquettes each of which share a site with two others (see  \figref{fig:single_clusters}(e)). 

There are compelling reasons for choosing such defect cluster configurations. First, is the energetics of the defect-defect interaction. As discussed in \smsec{SM:sec:SingleDefect}, the lowest energy configuration of two flux defects is when they are in the dipole configuration (see \figref{fig:single_clusters}(b)). It is therefore natural to expect flux disorder to appear as small defect clusters like the ones introduced in the previous paragraph. A second crucial reason for the choice of such defect clusters arises from other physical considerations relating to the nature of the zero-energy state. It is typical for systems with chiral symmetry to host zero modes around defects~\cite{Pereira2006,Ovdat2020,Mesaros2013}. Quite interestingly, each of the defect clusters discussed above hosts {\em distinct} types of zero energy modes, i.~e, the zero energy wavefunction decays with different power laws as discussed in \smsec{SM:sec:SingleDefect}. In the presence of a random ensemble of such defect clusters, one might expect that the zero-energy state obtained will arise from the hybridization of the zero modes localized around different defect clusters, and hence can have different characteristics for different clusters.

The central question we pose is the nature of fermionic wave functions at zero energy in the presence of a concentration $\conc$ of flux defects appearing in different types of defect configurations as shown \figref{fig:single_clusters}. It is important to point out that even in the presence of such flux disorder with differently correlated defect clusters, the system at half filling of fermions always is in the BDI symmetry class -- the same as the clean system.
Below we shall subject this model to a variety of numerical approaches, including exact diagonalization, and transfer matrix methods to investigate the nature of the states, and transport calculations which provide the nature of responses of the system. 


\section{Numerical Results
} \label{sec:NumericalResults}

In this section, we present the main results of various calculations, relegating details to \smsec{SM:sec:NumericalDetails}. 

\subsection{Transfer Matrix}\label{sec:NR:TM}

\begin{figure}
    \centering
    \includegraphics[width=\linewidth]{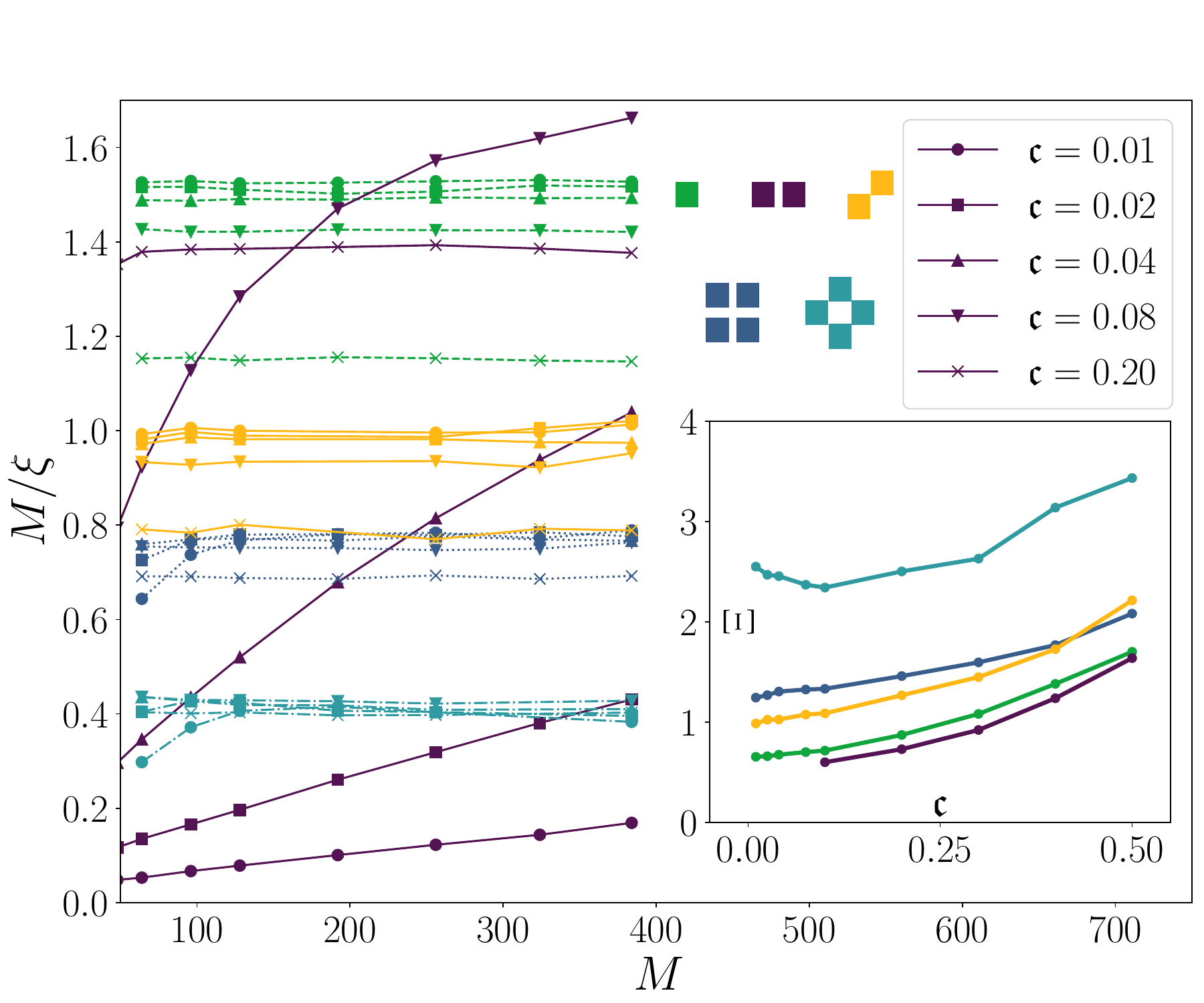}
    \caption{{\bf Results of Transfer Matrix Calculations:} The quantity $M/\xi$ as a function of $M$, where $\xi$ is a length scale associated with the zero energy state, $M$ is the width of the strip geometry. Different colors, as indicated in the legend, show results for different defect clusters, while different symbols indicate different concentrations $\conc$ flux defects. The inset shows the values (only well-converged values are shown) of $\Xi = \lim_{M \to \infty} \frac{\xi}{M}$ for various defect clusters.}
    \label{fig:tm_summary}
\end{figure}

The well-established transfer matrix method~\cite{markos2006numerical,ChalkerAPS1993scattering} allows for the extraction of a length scale $\xi$ associated with the wavefunction at a desired energy (zero energy in the present case). This scale $\xi$ is calculated in a strip of width $M~ (64 \leq M \leq 384)$ along the $2$-direction with periodic boundary conditions along $2$-direction, and lengths \( L \approx 2 \times 10^5 \) along the $1$-direction. The smallest eigenvalue of the transfer matrix scales as $e^{-\xi^{-1} L}$ where $\xi$ is a correlation (localization) length.
The quantity of interest is $M/\xi$: an increasing $M/\xi$ with increasing $M$ signals a localized state, while a decreasing $M/\xi$ with increasing $M$ indicates a delocalized state. The saturation of $M/\xi$ to a constant value indicates a critical state.

Fig.~\ref{fig:tm_summary} shows the results of the transfer matrix calculations for various defect cluster distributions and various concentrations. For all defect clusters considered here, $M/\xi$ attains a constant value at large $M$; the inset in \figref{fig:tm_summary} shows limiting values $\Xi = \lim_{M \to \infty} \frac{\xi}{M}$. 
The apparent exception is the case of dipole clusters at small concentration $\conc \lesssim 0.1$: we show in \smsec{SM:sec:ND:TM}, even in these cases the $M/\xi$ saturates to a constant value even if a reliable extraction of the quantity is numerically prohibitive.  The key qualitative inference obtained from these results is that the system in the thermodynamic limit attains {\em a critical (scale-invariant) state at zero energy}. The properties of the critical state are determined {\em both} by the defect concentration $\conc$ {\em and} the type of defect cluster - notably, two systems with the same $\conc$ realizing different defect clusters can have different values of $\Xi$.

\subsection{Transport Calculations}\label{sec:NR:Transport}

\begin{figure}
    \centering
    \includegraphics[width=\linewidth]{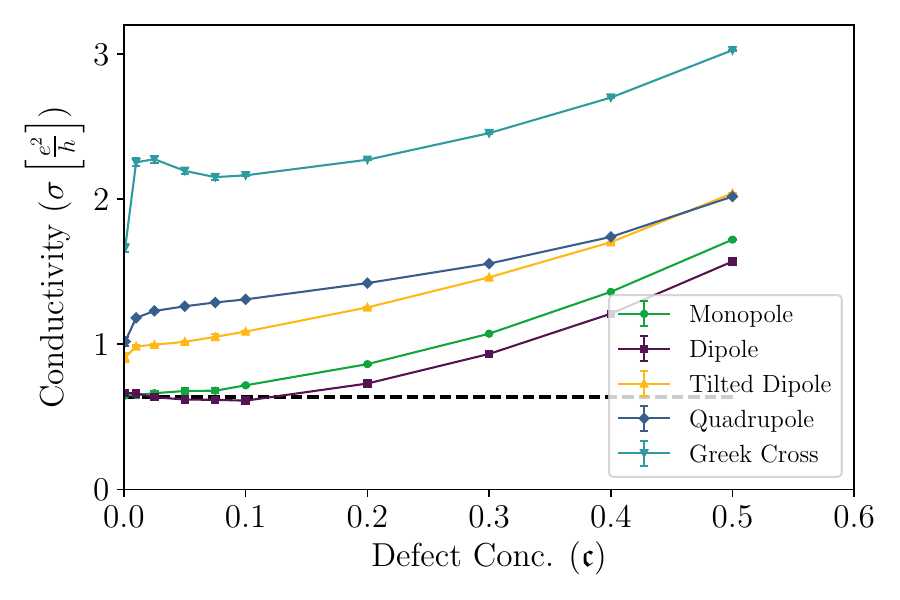}
    \caption{{\bf Conductivity Results:} Conductivity obtained from strip geometry with aspect ratio $4$ for different types of defect clusters. The dashed line shows $\sigma = \frac{2}{\pi} \frac{e^2}{h}$ for the clean system.}
    \label{fig:conductance_summary}
\end{figure}

An immensely useful quantity to characterize the state realized in the disordered system is conductivity $\sigma$. We calculate the conductivity by calculating the transmission probability across a strip geometry of length $L$ and width $M$ using an aspect ratio $M/L$ bigger than unity. Employing the {\tt Kwant}~\cite{Kwant2014} package for this calculation, we used an aspect ratio of $4$ and $W$ up to $512$. 

We recall that the clean system (\( \conc = 0 \)) exhibits a conductivity \( \sigma = \frac{2}{\pi} \frac{e^2}{h} \approx 0.637 \, \frac{e^2}{h} \), as established in Refs.~\cite{Ludwig1994,Ostrovsky2006,Schuessler2009}, and shown as the dashed line in Fig.~\ref{fig:conductance_summary}.
The said figure also shows the conductivity as a function of the concentration for different defect clusters (further details about conductivity calculations, including effect of aspect ratio, system size effects, statistical properties of conductance, etc., can be found in \smsec{SM:sec:ND:Transport}). Several features may be noted. In all cases, the conductivity approaches the value of the clean system as concentration $\conc \to 0$. However, the behaviour at finite concentrations is drastically different for different types of defect clusters. In the case of monopole and dipole defect distributions, $\sigma$ increases slowly from the clean value. On the other hand, for the other types of defect clusters, the conductivity rapidly increases from the clean limit with the increase of $\conc$ and then has a slower increase with further increase of $\conc$. Apart from the case of the Greek cross defect clusters, the $\sigma$ increases monotonically with increasing $\conc$. A notable feature is that in all cases the conductivity is of order $e^2/h$ even at concentration $\conc = 0.5$.

\subsection{Properties of Zero Energy States}\label{sec:IPR}

We employ numerical exact diagonalization to study the statistical properties of the zero-energy states. A key quantity of interest is the generalized inverse participation ratio (IPR) of a normalized state $\psi_{Ia}$ is
\begin{equation}\label{eqn:IPRdef}
    \IPR_q =  \sum_{I a} |\psi_{Ia}|^{2q}
\end{equation}
where the sum runs over the unit-cells ($I$) and flavor labels $(a)$ in the system, and $q$ is a real number. The quantity scales with linear system size $L$ as
\begin{equation}\label{eqn:taudef}
    \IPR_q \sim L^{-\tau_q}
\end{equation}
where the exponent $\tau_q$ provides information about the nature of the states~\cite{EversMirlin2008}.  

\begin{figure}
    \centering
    \includegraphics[width=1\columnwidth]{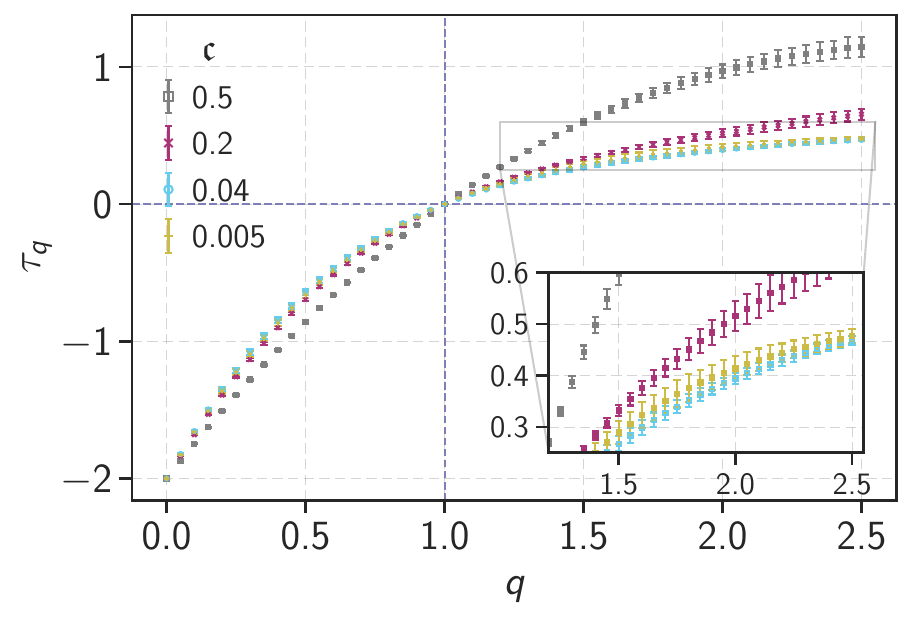}
    \caption{{\bf Multifractality of Zero Energy States:} Average $\tau(q)$ vs. $q$ for different $\conc$ for monopole defects.  
    }
    \label{fig:all_tau_avg}
\end{figure}

Fig.~\ref{fig:all_tau_avg} shows the dependence of $\tau_q$ as a function of $q$ for various concentrations $\conc$ for monopole disorder using system sizes as large as $2048\times2048$. The key feature to note is the multifractal nature of the zero-energy state, that does {\em not} show freezing~\cite{EversMirlin2008,Hafner2014} behaviour $\tau_q \to 0, q \to \infty$, but rather has termination~\cite{EversMirlin2008} character, i.~e., $\tau_q$ tends to a finite non-zero value as $q \to \infty$. We have confirmed this (see \smsec{SM:sec:ND:WFP}) by a study of the statistics of $\IPR_q$. 
We have also explored other defect clusters with smaller system sizes, and the results are qualitatively similar. The summary of the study of wavefunction statistics is that for all the defect clusters studied, we find that there is a multifractal termination behaviour when $\conc \gtrsim 0.1$, for smaller defect concentrations we are unable to rule out freezing with the available computational resources.

\section{Discussion}\label{sec:Theory}

This section explores ways of understanding the numerical results presented in the previous section from field theoretical considerations. The first step towards this is to find a long-wavelength description of the gauge field disorder introduced by the flux defects. To this end, we write the disordered Hamiltonian as
\beq\label{eqn:HamUV}
H = H_0 + H_{F}
\eeq
where $H_0$ is the Hamiltonian of the clean system with the hoppings as described in \eqnref{fig:schematic_of_model}, and 
\beq\label{eeqn:HamF}
H_F = -t\sum_{I \delta} \Delta z_{ab}(I,\delta)c^\dagger_{(I+\delta)a}c_{Ib}
\eeq
where $\Delta z_{ab}(I,\delta) = -2 z_{ab}(I,\delta)$ if the gauge field on the link is flipped, and zero otherwise. To obtain a continuum field theoretical description of the disordered system, after noting that $H_0$ in the continuum limit is described by the Dirac Hamiltonian ${\cal H}_0$ given in \eqnref{eqn:Dirac0}, $H_F$ can be coarse-grained as
\beq\label{eqn:DiracF}
\begin{split}
{\cal H}_F= &  \int \D{^2 \br} \Psi^\dagger(\br) \bW(\br) \Psi(\br) \\
 & + \int \D{^2 \br} \Psi^\dagger(\br) 
 \left[ \sum_i \left(\bV_i(\br) + \bV^\dagger_i(\br) \right)  \right]\Psi(\br) \\
  + \int \D{^2 \br} & \left( \dou_i \Psi^\dagger(\br) \bV_i(\br) \Psi(\br) + \Psi^\dagger(\br) \bV^\dagger_i(\br) \dou_i \Psi(\br)\right) \Psi(\br)
\end{split}
\eeq
where matrices $\bW(\br)$ are determined by the changes of the gauge fields in the unit-cell located at $\br$ and $\bV_i(\br)$ arise from the change of the gauge fields along the links that connect unit cell at $\br$ and that at $\br+ e_i$ $i\in \{1,2\}$. See \smsec{SM:sec:FTD} for details.

\begin{figure}
    \centerline{\includegraphics[width=0.5\linewidth]{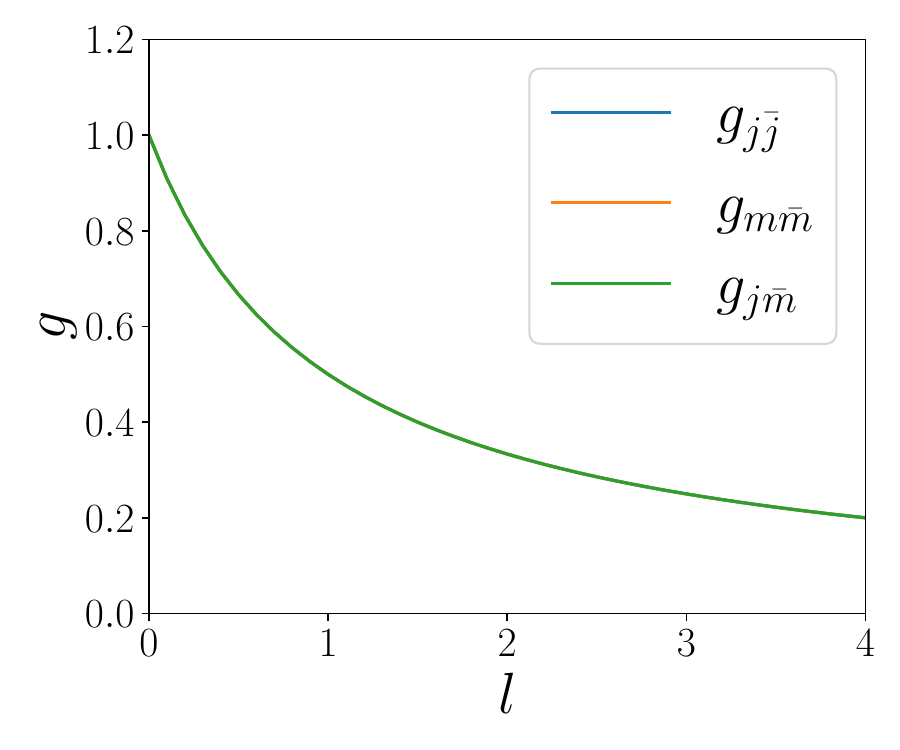}
    \includegraphics[width=0.5\linewidth]{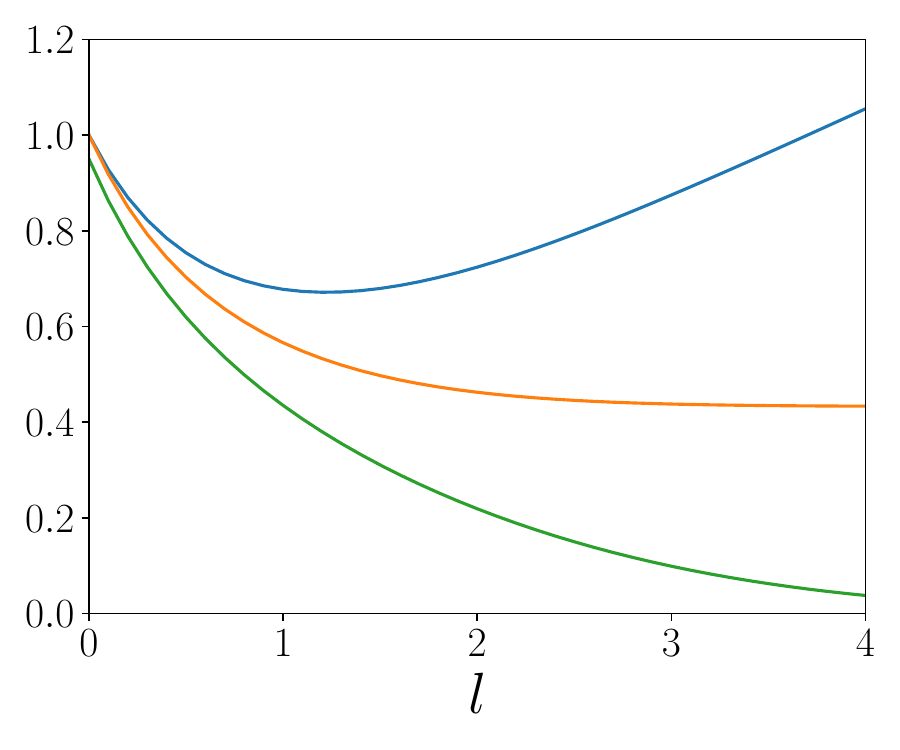}}
\centerline{~~~~(a) ~~~~~~~~~~~~~~~~~~~~~~~~~~~~~~~~~~ (b)}
    \caption{{\bf RG flow results: }(a) In presence of perfect mass-gauge correlation i.e. $g_{j\bar{j}} = g_{m\bar{m}} = g_{j\bar{m}} = 1$ all the coupling parameter decays to zero under RG flow. (b) For less correlated values ($g_{j\bar{j}} = g_{m\bar{m}} =0;  g_{j\bar{m}} = 0.95$), the disorder strength $g_{j\bar{j}}$ initially decreases but eventually goes to infinity.  }
    \label{fig:RGresults}
\end{figure}

Before we discuss the present case of flux disorder, we will briefly recall~\cite{HatsugaiWen1997,Guruswamy2000} the analysis of disordered Dirac fermions in the BDI class. By a unitary transformation of the spinor fields $\Psi(\br)$ (see \eqnref{eqn:Dirac0}) the matrices $\alpha_i$ are transformed to $\sigma_1 \otimes \sigma_i$, and 
\beq\label{eqn:DiracStandardForm}
\begin{split}
{\cal H} & = {\cal H}_0 + {\cal H}_F = \int \D{^2 \br} \Psi^\dagger(\br) \begin{pmatrix} 0 & {\cal D} \\
{\cal D}^\dagger & 0 
\end{pmatrix} \Psi(\br)\\
&{\cal D} = \sigma_i (-\ci \dou_i) + \sigma_i A_i(\br) + \tau_i m_i(\br) 
\end{split}
\eeq
where the gauge fields $A_i(\br)$ and mass terms $m_i(\br)$ are random real fields capturing the disorder in BDI symmetry class, $\tau_1 = \sigma_0, \tau_2 = \sigma_3$. Note that the gradient terms as usually neglected as irrelevant. In previous works~\cite{HatsugaiWen1997,Guruswamy2000,Mudry2003}, these random fields are taken as zero-mean $\delta$-correlated random fields, with the variance of $A_i$ described by a positive real number $g_{j\bar{j}}$, and that of $m_i$ described by $g_{m\bar{m}}$. When $g_{m\bar{m}}=0$, $g_{j\bar{j}}$ does not flow (under renormalization group) and one obtains a multifractal state with a conductance given by $\frac{2}{\pi} \frac{e^2}{h}$~\cite{Ludwig1994,Guruswamy2000}. On the other hand, when $g_{m\bar{m}} \ne 0$, $g_{m\bar{m}}$ flows to a finite value under renormalization group transformations, and $g_{j\bar{j}} \to \infty$, leading to strong coupling. The sigma-model approach of Gade-Wegner~\cite{Gade1991,Gade1993,Guruswamy2000,Motrunich2002,Mudry2003} then shows that the resulting state has a characteristically divergent density of states, and a conductivity unchanged from the clean system~\cite{Ludwig1994}. Furthermore, the results of ref.~\cite{Chamon1996} suggest that at large $g_{j\bar{j}}$, the zero energy states have frozen multifractality.

\begin{figure}
    \centering
    \includegraphics[width=\linewidth]{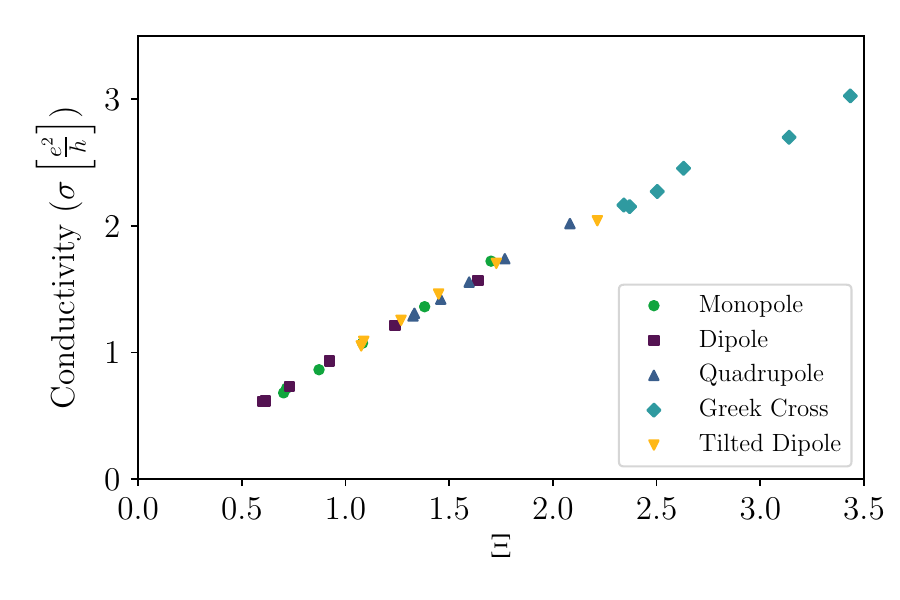}
    \caption{{\bf A family of critical points:} Conductivity vs $\Xi$ for various defect clusters and concentrations. Only well-converged data for $\conc \gtrsim 0.05$ data are used to obtain this correlation.}
    \label{fig:universality}
\end{figure}

A simple-minded application of the results reviewed in the previous paragraph to the case of flux disorder via the field theory \eqnref{eqn:DiracF}, does not help explain the results obtained in the previous section. For example, flux disorder introduces terms akin to gauge and mass disorder; however, for $\conc \gtrsim 0.1$, we do not find terminating multifractality, and conductivity is significantly and systematically different from the clean system. A closer look reveals several points to be taken into account. First, the disorder in \eqnref{eqn:DiracF} is {\em not zero mean}, in fact, the mean disorder can be shown to lead to the change of the Dirac velocity by a factor proportional to $\conc$ (see \smsec{SM:sec:RG}). This, however, cannot produce the qualitative differences that we find. Second, we find that gauge and mass disorder introduced by flux disorder are {\em correlated}, i.~e., the $A$ and $m$ fields in \eqnref{eqn:DiracStandardForm} are correlated. We have performed renormalization group (see \smsec{SM:sec:RG}) calculations~\cite{Efetov1996,Guruswamy2000} where we have introduced $g_{j\bar{m}}$ which describes the correlations between  random fields $A_i$ and $m_i$, resulting in flow equations 
\begin{equation}
\begin{split}
        \frac{\D g_{j\bar{j}}}{\D l} & = g_{m\bar{m}}^2 - 2|g_{j\bar{m}}|^2 \\
        \frac{\D g_{m\bar{m}}}{\D l} & = - |g_{j\bar{m}}|^2 \\
        \frac{\D g_{j\bar{m}}}{\D l} & =  - g_{j\bar{j}}g_{j\bar{m}}
        \label{eqn:rgflowend}
\end{split}
\end{equation}

The central outcome of this study is that in the presence of $g_{j\bar{m}}$, $g_{j\bar{j}}$ flows to {\em smaller} values. Only above a certain length scale (when the magnitude of $g_{j\bar{m}}$ has gone to a small value) does $g_{j\bar{j}}$ goes to larger strong coupling values while $g_{m \bar{m}}$ similar to the results quoted in the previous paragraph when $g_{j\bar{j}}$ and $g_{m\bar{m}}$ are uncorrelated. This analysis, therefore, suggests the possibility that the non-freezing terminating multifractality that we find is a consequence of the fact the systems sizes that we have studied are below the length scale where the correlations between the gauge and mass disorder become small. While we cannot rule out this scenario, in the next paragraph, we offer compelling evidence that this is not the case and that there is entirely new physics at play here.

The transfer matrix result strongly suggests a scale invariant state characterized by $\Xi$ which depends both on the type of defect clusters, as well as, on the concentration $\conc$. For all types of defect clusters, we are able to obtain well-converged values of $\Xi$ and the conductivity in the thermodynamic limit when $\conc \gtrsim 0.05$. It is natural to enquire if $\Xi$ is related to the conductivity $\sigma$. Fig.~\ref{fig:universality} shows the relationship between $\sigma$ and $\Xi$, for different types of defect clusters and for concentrations $\conc \gtrsim 0.05$. Quite remarkably, we see that these data fall on a ``universal'' line! These results suggest that the presence of flux defects produces a continuous family of critical states with certain universal properties that vary continuously along the critical line. We have not been able to analytically access this critical manifold by perturbing the Dirac point. Indeed, gauge disorder induced by flux defects is non-perturbative and spatially correlated, and an analytical theory for this is likely challenging.

\begin{figure}
    \centering
    \includegraphics[width=1.0\columnwidth]{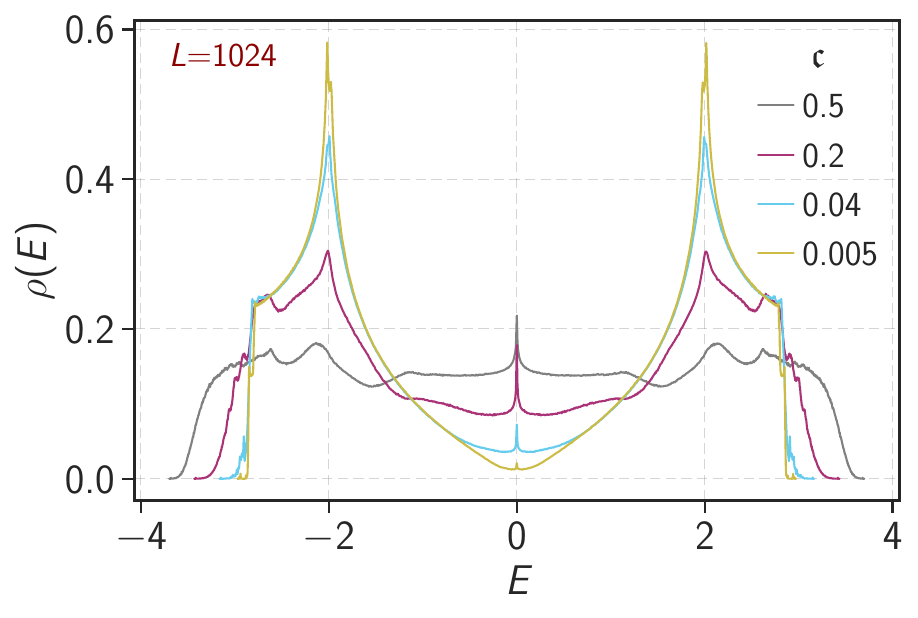}
    \caption{{\bf DOS:} Density of state for a square sample of linear size $L=1024$ for different values of $c={0.5, 0.2, 0.04, 0.005}$ for monopole defects showing the diverging behavior at the band-center. 
    }
    \label{fig:dos}
\end{figure}

\begin{figure}
    \centering
    \includegraphics[width=\linewidth]{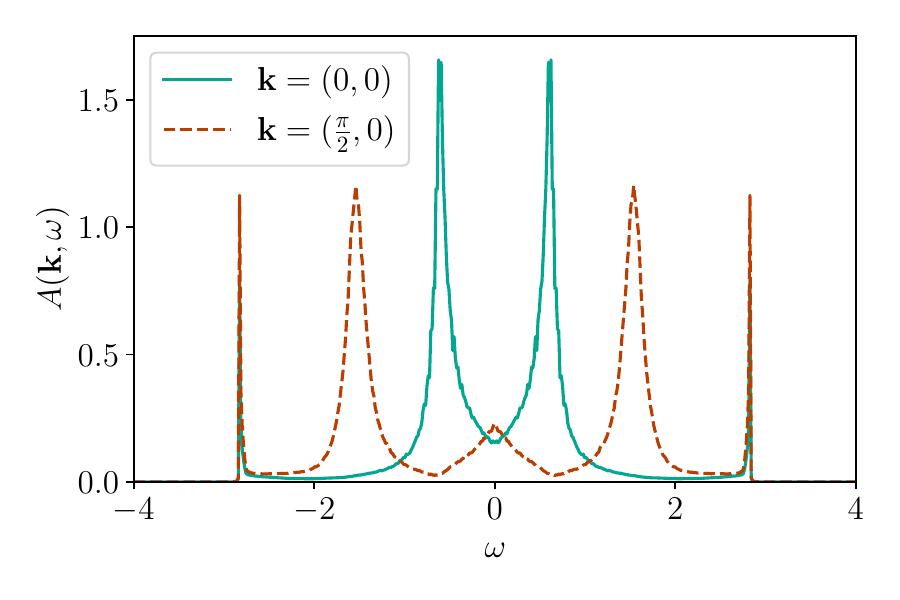}
    \caption{{\bf Spectral Function:} Numerically obtained spectral function by disorder-averaging an ensemble of Greek cross defect clusters with concentration $\conc = 0.1$ in $128\times128$ system. $128$ samples are used for disorder averaging.  }
    \label{fig:numerical_spectral_function}
\end{figure}

It is natural to enquire if these new critical points are also characterized by a divergent density of states. Fig.~\ref{fig:dos} shows that there is indeed a sharp feature about a background in fashion similar to the Gade-Wegner singularity. However, the numerical resources available to us do not allow us to study these features to obtain a quantitative understanding. Nevertheless, we have attempted to explore how the sharp feature in the density of states arises, i.~e, how the states of the clean system reorganize in the presence of flux disorder to produce the sharp feature in the density of states. To this end, we numerically calculated the disorder-averaged spectral function~\cite{Joao2022} associated 
\beq
\begin{split}
{\cal G}_{ab}(\bk,z) & = \disave{\bra{\bk,a}(z - H)^{-1} \ket{\bk,b}}, \\
A(\bk,\omega) & = -\lim_{\eta \to 0^+}\frac{1}{\pi} \Im\left(\sum_{a} {\cal G}_{aa}(\bk,\omega + \ci \eta)\right) 
\end{split}
\eeq
with Bloch states at a point $\bk$ in the Brillouin zone, $a$ is the flavor label, $z$ is a complex frequency, $H$ is the Hamiltonian \eqnref{eqn:HamUV}, and $\omega$ is a real frequency, and $\disave{~}$ is the disorder average. Fig.~\ref{fig:numerical_spectral_function} shows the spectral function $A(\bk,\omega)$ for two $\bk$ points in the Brilloun zone for Greek cross defect-cluster at concentration $\conc = 0.1$. Quite interestingly, the states at the Dirac point are pushed away from the zero energy (blue curve), while states significantly away from the Dirac point are pushed towards zero energy! This calculation also emphatically demonstrates that the new class of critical points discovered here cannot be accessed perturbatively from the clean system Dirac state. We have attempted to understand the underlying physics by a non-perturbative $T$-matrix calculation (see \smsec{SM:sec:Tmatrix}). A key finding of that section is that {\em coherent scattering between defect clusters} is key to producing this effect of spectral redistribution. 






\section{Summary and Outlook}\label{sec:Summary}

In this paper, we have studied the physics of fermions on a square lattice with $\pi$-flux experiencing $\Integers_2$ gauge field disorder induced by flux defects (plaquettes with zero flux). A central finding is that the system goes to a new type of non-perturbative critical state whose properties depend not only on the defect concentration, but also on the type of defect clusters. The most interesting aspect is that these critical states appear to belong to a family of non-perturbative fixed points characterized by a conductivity of order $e^2/h$ that is related to the Lyapunov exponent obtained from the transfer matrix calculations.  

These results offer a rich phenomenology for the physics of fermions moving in the background of discrete gauge fields, as for example studied in various contexts~\cite{Kitaev2006,Gazit1}. At low temperatures, one might expect defects to appear in clusters at concentrations determined by the energetics. Further, at larger temperatures, higher concentrations of defects will result, but with smaller clusters. The result uncovered in \figref{fig:universality} (note that thermal effects of fermion occupation of states are not included in that figure) may be useful in understanding the temperature-dependent transport phenomena. It may be interesting to explore these ideas in material systems~\cite{Hermanns2018,Trebst2022}. 

We conclude the paper by noting that an analytical description of the non-perturbative family of disordered critical points uncovered in this work offers an interesting new research direction. Another interesting direction is to investigate the honeycomb lattice~\cite{Zhuang2023} to explore if a similar family of critical states is realized there.



\noindent
{\bf Acknowledgements:} HD acknowledges support from Ministry of Education via a PMRF Grant. VBS is supported by DST-SERB. S.B. would like to thank MPG for funding through the Max Planck Partner Group at IITB. N.N. would like to thank DST-INSPIRE fellowship, Grant No. IF- 190078, for funding. We thank Ferdinand Evers, Subhro Bhattacharjee, Diptiman Sen, and Sumilan Banerjee for discussions. 

\bibliography{ref}

\clearpage
\newpage

\appendix
\renewcommand{\thefigure}{\thesection.\arabic{figure}}
\renewcommand{\theequation}{\thesection.\arabic{equation}}
\counterwithin{figure}{section}

\section{Single Defect Results}\label{SM:sec:SingleDefect}

\begin{figure}
    \centering
    \centerline{\includegraphics[width=\linewidth]{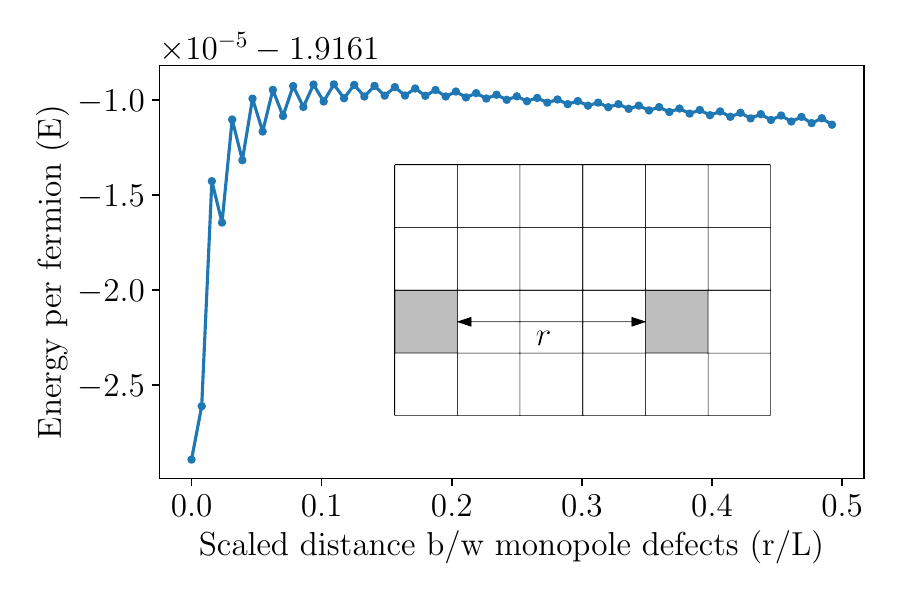}}
    \caption{{\bf Defect stability: }Variation of energy per fermion (in the unit of hopping amplitude $t$) as distance $r$ between the $\pi$ flux defects increased in a system of size $128\times128$. The most stable configuration is $r=0$ i.e. a dipole defect cluster.}
    \label{SM:fig:interaction}
\end{figure}

In the free $\mathbb{Z}_2$ gauge theory, the $\mathbb{Z}_2$ flux defects don't interact with each other as the energy of the system doesn't change due to change in the separation distance between the flux defects. However, coupling the $\mathbb{Z}_2$ gauge fields with fermions leads to an effective interaction between the flux defects. This is illustrated by considering a clean system at half-filling and introducing two $\pi$ flux defects at variable distance ($r$). We find that the energy of the system is minimized when the defects are adjacent to each other forming the dipole defect cluster as described in the main text.  \figref{SM:fig:interaction} shows the variation of the energy per fermions as a function of distance between the two defects. This calculation suggests that, at the low temperature, the system will contain mostly dipole defect cluster. 

\begin{figure*}
    \centering
    \centerline{
    \includegraphics[width=0.33\textwidth]{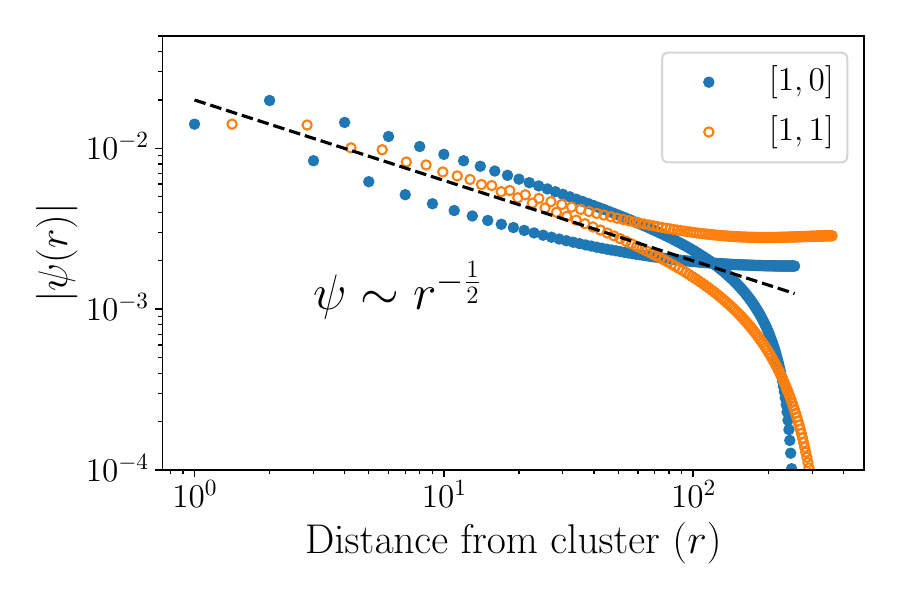}
    \includegraphics[width=0.33\linewidth]{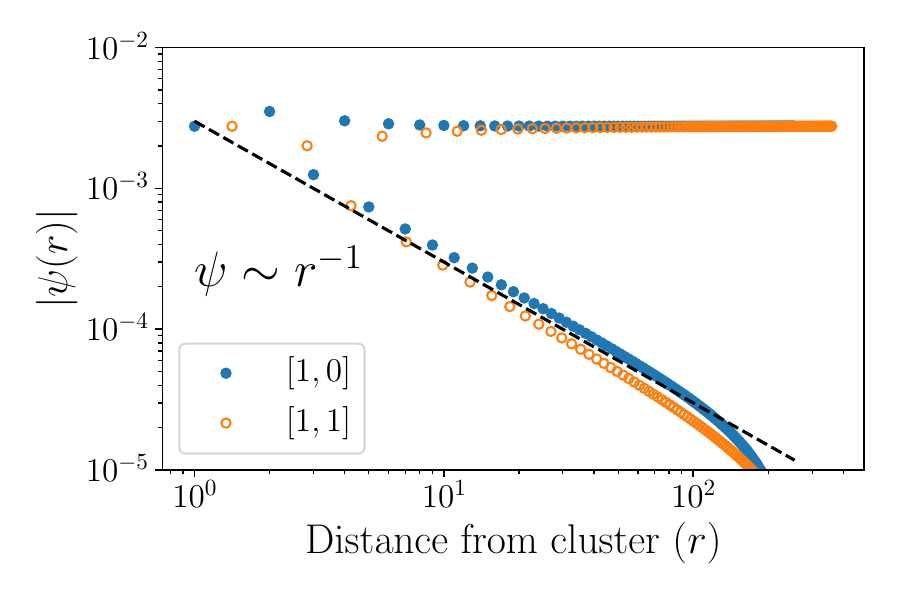}
    \includegraphics[width=0.33\linewidth]{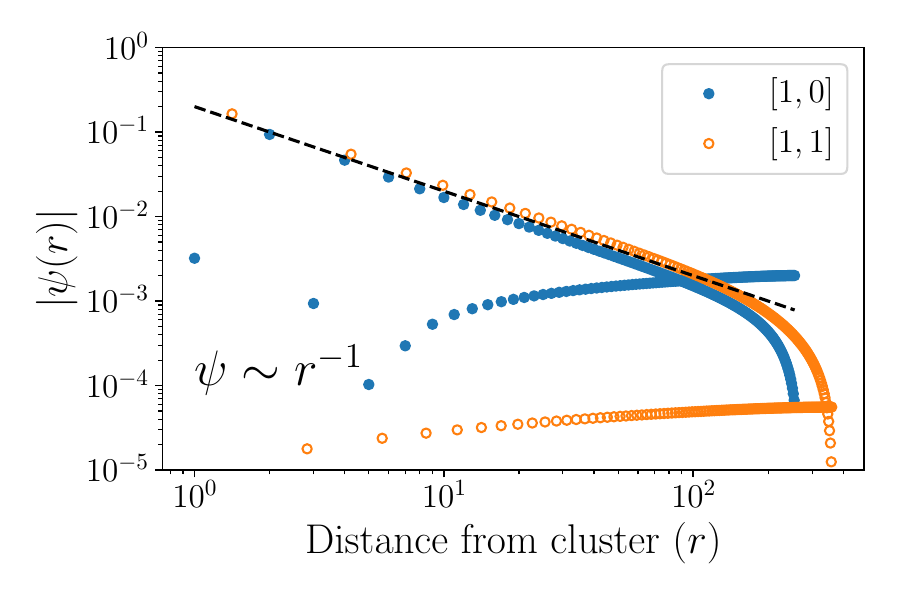}
    }
    \centerline{ ~~~~~~~~~~~~~~~~~~~~~~~~~~~~~~~~~~~~~~~~(a)~~~~~~~~~~~~~~~~~~~~~~~~~~~~~~~~~~~~~~~~~~~~~~~~~~~(b)~~~~~~~~~~~~~~~~~~~~~~~~~~~~~~~~~~~~~~~~~~~~~~~~~~~~~(c)~~~~~~~~~~~~~~~~~~~~~~~~~~~~~~~~~~~~~~~~}
    \centerline{
    \includegraphics[width=0.33\linewidth]{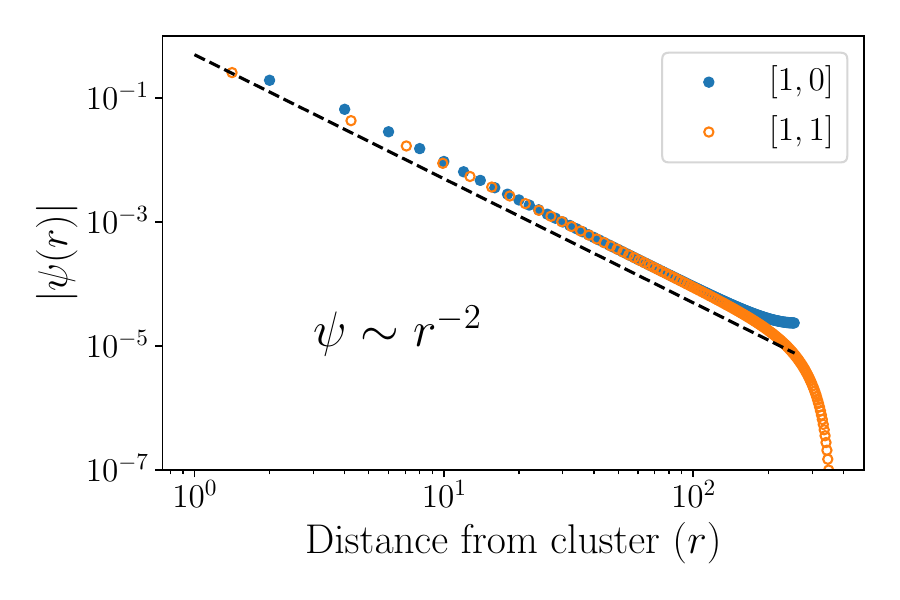}
    \includegraphics[width=0.33\linewidth]{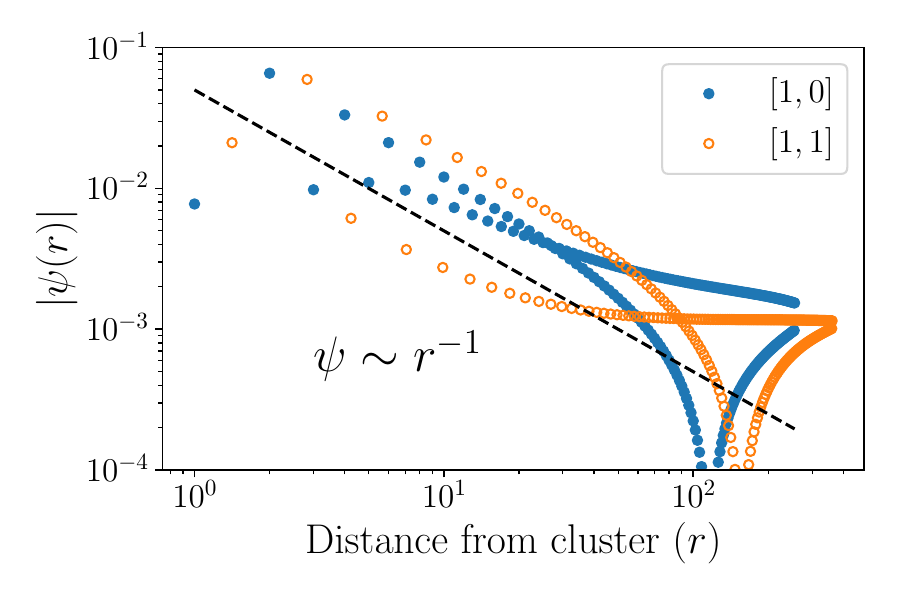}
    }
    \centerline{ ~~~~~~~~~~~~~~~~~~~~~~~~~~~~~~~~~~~~~~~~~~~(d)~~~~~~~~~~~~~~~~~~~~~~~~~~~~~~~~~~~~~~~~~~~~~~~~~~~~~(f)~~~~~~~~~~~~~~~~~~~~~~~~~~~~~~~~~~~~~~~~}
    \caption{{\bf Power law decay of wavefunction: }Power law decay of wavefunctions around the $\pi$ flux defect cluster along the crystallographic direction $[0, 1]$ and $[1,1]$. The defects shown in the graphs are (a) monopole (b) dipole, (c) tilted dipole, (d) quadrupole, and (e) Greek cross. The dashed lines are used as reference lines for the power law behavior.}
    \label{fig:powerlaw}
\end{figure*}

Different defect clusters are found to produce different types of zero-energy states characterized by the power-law decay of the wavefunction from the position of the defect cluster. To study the power-law decay of the zero-energy wavefunction around the defect, we place the defects at the center of a system of size $512\times512$ and compute the states with energy closest to zero. The absolute value of the wave function along the positive $x$ direction and along the diagonal direction is summarized in the plots shown in \figref{fig:powerlaw}. We find that the zero-energy wavefunction decays as  $r^{-1/2}$ for monopole defects whereas for quadrupole defect, it decays as $r^{-2}$. Such difference in the properties of zero energy wavefunctions motivated us the study the transport properties of the defects separately. It must be noticed that the zero energy wavefunction due to cluster defects in some cases also saturates to nonzero values as the distance from the defect cluster is increased. This suggests that the effect of a single defect is non-local and they can hybridize significantly with a zero energy wavefunction of a distant defect cluster.

\section{Numerical Details}\label{SM:sec:NumericalDetails}
\subsection{Transfer Matrix}\label{SM:sec:ND:TM}

\begin{figure}
    \centerline{
    \includegraphics[width=0.5\linewidth]{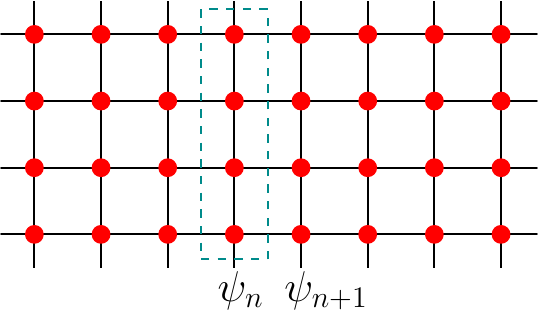}
    \includegraphics[width=0.5\linewidth]{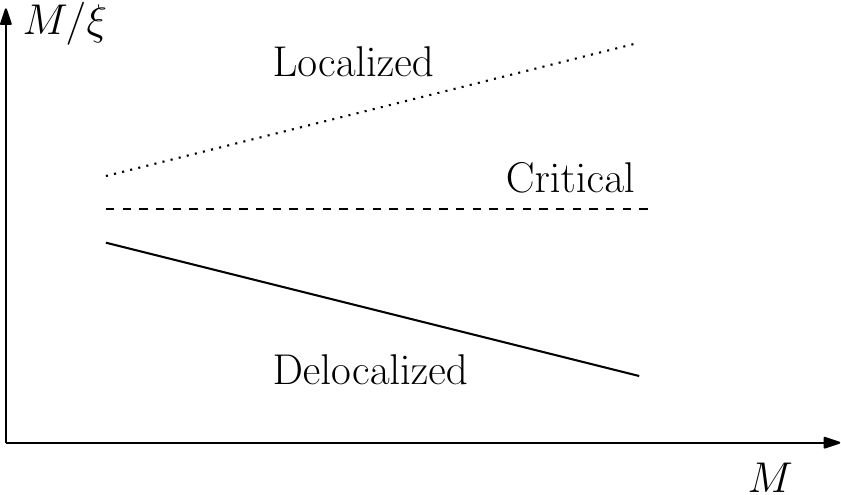}
    }
    \centerline{(a)~~~~~~~~~~~~~~~~~~~~~~~~~~~~~~~~~~~~~(b)}
    \caption{{\bf Transfer Matrix Schematic: } (a) Showing the slices of the system used in the transfer matrix calculation (b) Interpretations of transfer matrix results.}
    \label{SM:fig:tmbasics}
\end{figure}

We use transfer matrix method to find the localization length along the $x$ axis of our systems by choosing a quasi one dimensional geometry with \( L = 2 \times 10^5 \) and $64 \leq M \leq 384$. We divide the system into layers of one dimensional system and denote the wavefunction amplitude of $n$-th layer by $\psi_n$ as shown in the \figref{SM:fig:tmbasics} (a). The wavefunction amplitude corresponding to $(n+1)$-th layer is related to the same of $n$-th and $(n-1)$-th layer as
\begin{equation}
    \begin{pmatrix}
        \psi_{n+1} \\ 
        \psi_{n}
    \end{pmatrix} = \begin{pmatrix}
        T_n^{-1}\left(E - H\right) & -T_n^{-1}T_{n-1} \\ 
        \mathbbm{I} & 0
    \end{pmatrix} 
    \begin{pmatrix}
        \psi_{n} \\ 
        \psi_{n-1}
    \end{pmatrix}.
\end{equation}
The matrix in the above equation is called the transfer matrix and denoted by $M_n$. Wavefunctions corresponding to the first and the last slice of the system is related by 
\begin{equation}
    \begin{pmatrix}
        \psi_{L} & \psi_{L-1}
    \end{pmatrix}^T  = M^{(L)} 
    \begin{pmatrix}
        \psi_{1} & \psi_{0}
    \end{pmatrix}^T
\end{equation}
where $M^{(L)} = \prod_{n=0}^{L-1}M_n$. We calculate the smallest Lyapunov exponent $\zeta_L$ of $M^{(L)}$ which is reciprocal of the localization length ($\xi$)  along the longer direction. The ratio $\xi/M$ or $M/\xi$ as a function of system's width tells us about the nature of the states at energy $E$ as summarized in \figref{SM:fig:tmbasics} (b).

\subsubsection{Dipole Defects}

\begin{figure}
    \centering
    \includegraphics[width=\linewidth]{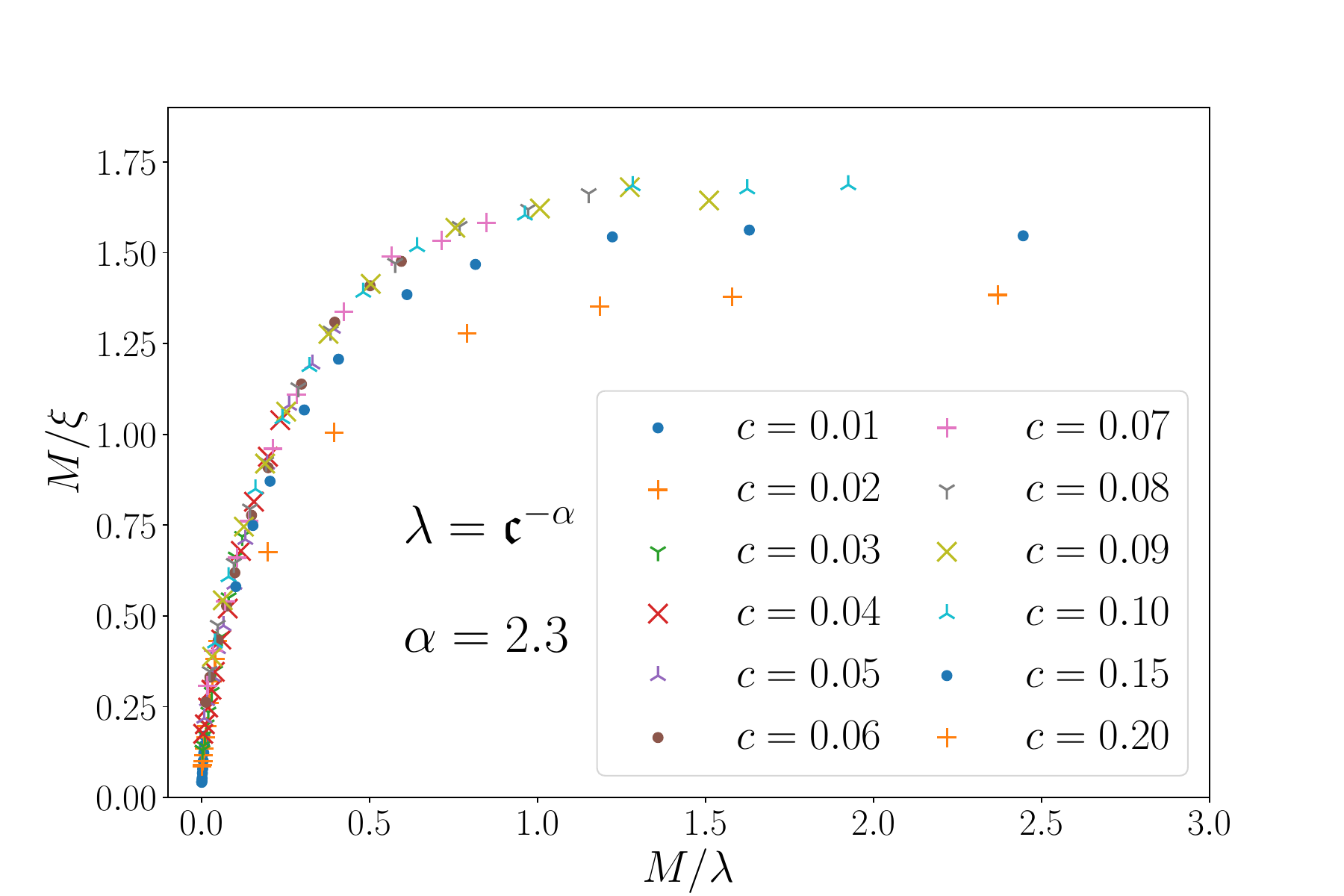}
    \caption{\textbf{Saturation of $M/\xi$ for dipole defect clusters: }Collapse of the $M/\xi$ data for small concentration $\conc \lesssim 0.1$ suggesting a length scale $\lambda \sim \conc^{-2.3}$ where the value of $M/\xi$  saturates.}
    \label{SM:fig:dipole_TM}
\end{figure}

From the transfer matrix calculation for $\pi$-flux defects we conclude that the zero energy state of the system is a critical state for all values of $c$. Although the result for dipole defects at low concentration of defects shows behavior similar to localized phase, here we justify that the zero-energy states corresponding to dipole defects are also critical states. We note that for higher values of defect concentration, e.g., $\conc = 0.2$, $M/\xi$ is a constant function of system's width $M$ and for $\conc = 0.1$, $M/\xi$ saturates as we increase $M$. We show using the \figref{SM:fig:dipole_TM} that for any values of $\conc$, there is a length scale where $M/\xi$ will saturate and the length scale depends on $\conc$ as $\conc^{-2.3}$.  

\subsection{Transport Calculations}\label{SM:sec:ND:Transport}
We have used {\tt Kwant} package to calculate the conductivities of our system in presence of different types of defects. {\tt Kwant} package calculates the transmission coefficients of the sample by attaching two semi-infinite leads to the opposite ends of the sample. Transmission coefficient $T$, calculated using {\tt Kwant}, is used to get the conductance of the system using Landauer-Büttiker formula
\begin{equation}
    G = \frac{e^2}{h}T.
\end{equation}
Conductivity of the system is related to the conductance as 
\begin{equation}
    \sigma = G\frac{L}{M},
\end{equation}
where the $L$ and $M$ are respectively the length and width of the system. In our calculation we have used a rectangular geometry where $M > L$ and we have shown below that our result doesn't change as we change the $M/L$ from $2$ to $4$.


\begin{figure}
    \centering
    \centerline{\includegraphics[width=\linewidth]{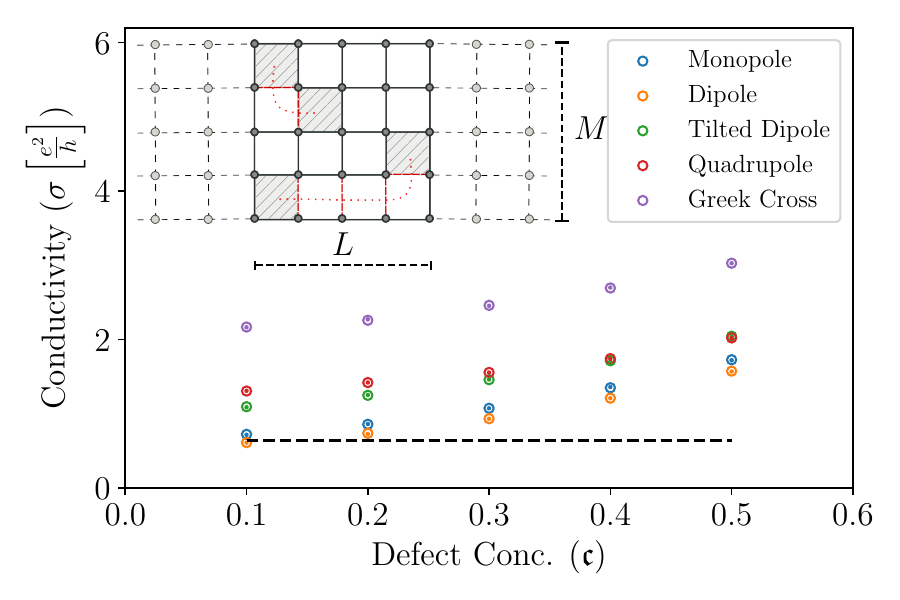}}
    \caption{{\bf Aspect ratio dependency: } Conductivity for each type of defect clusters is shown for different defect concentrations $\conc$. Data corresponding to $M/L = 2$ and $4$ are shown with circles and dots of same colors respectively. The width $M$ of the system for each calculation is $512$.}
    \label{SM:fig:cond_aspect_ratio}
\end{figure}
\subsubsection{Aspect Ratio Dependence}
One of the properties of the critical state is that the conductivity is independent of aspect ratio in the thermodynamic limit.
We have verified and shown in \figref{SM:fig:cond_aspect_ratio} that indeed the conductivity at half filling is independent of the aspect ratio of the system for a range of defect concentrations and for all the defect clusters. 

\begin{figure}
    \centering
    \includegraphics[width=1\columnwidth]{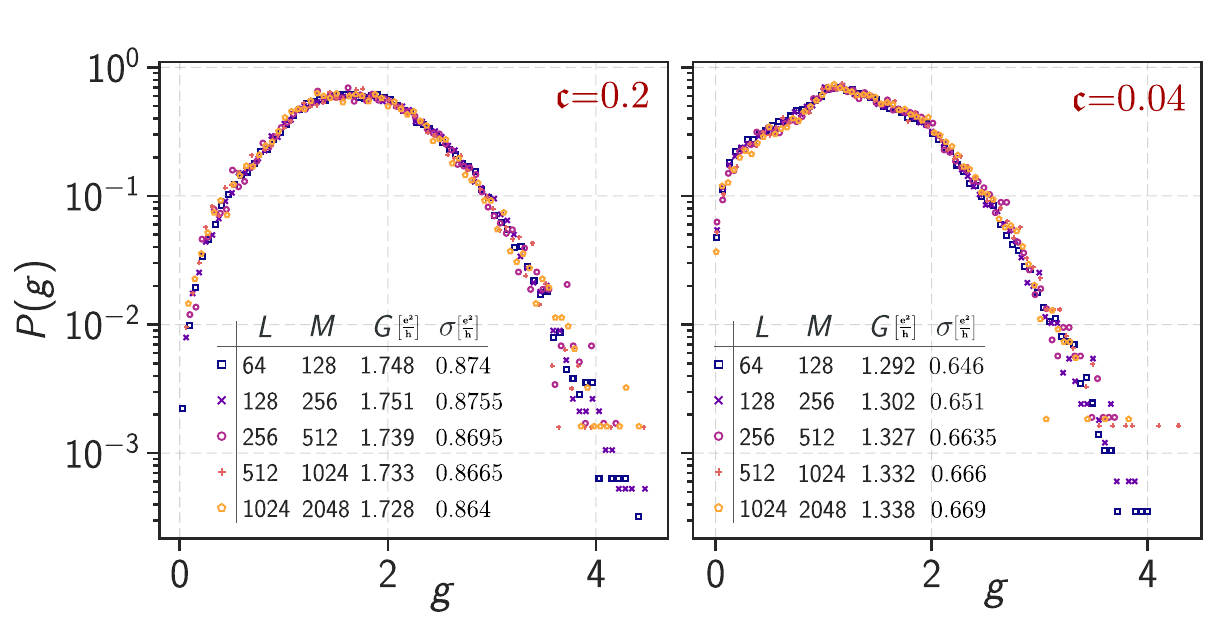}
    \caption{{\bf Conductance distribution: } Conductance distribution $P(g)$ for monopole defects for $\conc={0.2, 0.04}$ for aspect ratio $M/L=2$ at $E=0$. The scale invariance of the conductance distribution indicates the critical nature of the zero energy states.}
    \label{SM:fig:sigma_stat}
\end{figure}

\subsubsection{Conductance Distribution}
In the critical phase, the conductivity of the system should not change with system size. However this is true only for large enough system size. Moreover, the conductance distribution should also remain invariant as the system size is increased. To determine whether the systems we studied are sufficiently large, we have shown the conductance distributions for monopole defect clusters in \figref{SM:fig:sigma_stat}. For all other defect clusters except the Greek cross, we observe a well-converged conductivity distribution. In the case of the Greek cross cluster, the conductance distribution flows toward the higher conductance, suggesting an even larger conductance in the thermodynamic limit. 

\begin{figure*}
    \centering
    \includegraphics[width=0.95\textwidth]{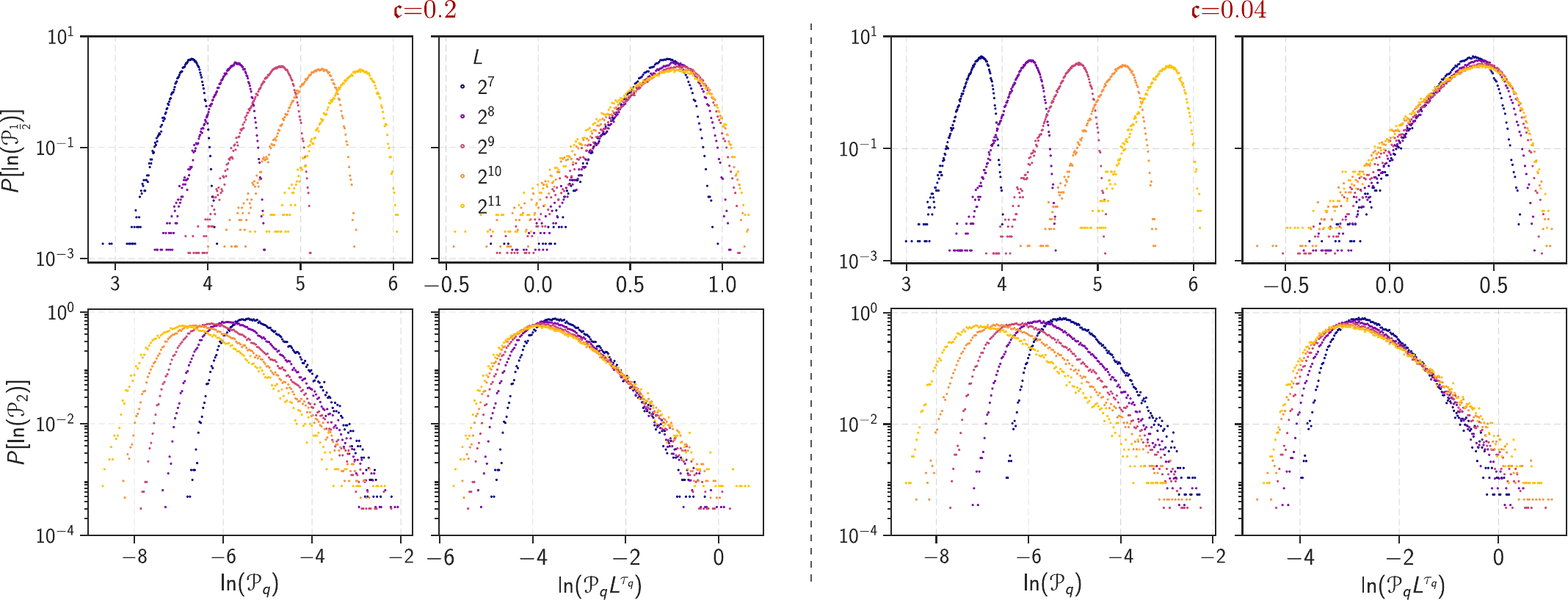}
    \caption{{\bf IPR distribution scaling:} Flow of the probability distribution of the inverse participation ratio $\IPR_q$ for two representative values of $q = 0.5$ and $q = 2$, shown across system sizes $L = 2^7, 2^8, \ldots, 2^{11}$, for monopole defect clusters at flux defect concentrations $\conc = 0.2$ and $\conc = 0.04$. The top panel displays the distribution $P(\IPR_{0.5})$, while the bottom panel shows $P(\IPR_2)$, each comparing the evolution of the distribution with increasing system size. The near invariance of the scaled distributions $\IPR_q L^{\tau_q}$ at large $\conc$ indicates scale-invariant multifractal statistics, while stronger finite-size effects are visible at small $\conc$. 
    }
    \label{SM:fig:IPRDist}
\end{figure*}
\subsection{Wavefunction Properties}\label{SM:sec:ND:WFP}
To analyze the statistical properties of the zero-energy states in our system, we perform exact diagonalization to compute the generalized inverse participation ratios (IPR), as defined in Eq.~\ref{eqn:IPRdef}. The IPRs reveal the degree of localization or delocalization of wavefunctions and allow for the extraction of multifractal scaling exponents via IPR scaling as in Eq.~\ref{eqn:taudef}. We consider system sizes up to $2048\times2048$, and explore a broad range of flux defect concentrations $\conc \in [0.005, 0.5]$, averaging over at least $10^3$ of disorder ensembles for each individual data points.

We find that the zero-energy states exhibit multifractal scaling, with clear indications of termination rather than freezing at large moments. Specifically, the scaling exponent $\tau_q$ does not saturate to zero as $q \to \infty$, which would indicate freezing, but rather levels off at a finite value, indicative of multifractal termination. This behavior is evident from both the scaling of $\tau_q$ curves and the distributions of $\ln(\IPR_q)$ shown in Fig.~\ref{fig:all_tau_avg} and \ref{SM:fig:IPRDist}, where we display results for both large and small values of $\conc$. Although finite-size effects are significant for small concentrations (e.g., $\conc \lesssim 0.05$), especially in differentiating the tail behavior of $\tau_q$, the data at larger concentrations demonstrate scale-invariant distributions, reinforcing the absence of wavefunction freezing. The absence of freezing across a range of $c$ and disorder types strongly supports the scenario that these zero-energy states form a new class of critical wavefunctions, distinct from both fully localized and strongly multifractal frozen states observed in other disordered Dirac systems~\cite{Hafner2014}.
\section{Field Theoretical Formulation of Flux Disorder}\label{SM:sec:FTD}

Next, we explore how this disorder affects the low-energy Dirac fermions. The long-wavelength theory is
\beq
\begin{split}
{\cal H}_0 = &  \int \D{^2 \br} \Psi^\dagger(\br) \alpha_i (-\ci \dou_i) \Psi(\br) \\
\mathcal{H}_F  = &  \int \D{^2 \br} \Psi^\dagger(\br) \bW(\br) \Psi(\br) \\
 & + \int \D{^2 \br} \Psi^\dagger(\br) 
 \left[ \sum_i \left(\bV_i(\br) + \bV^\dagger_i(\br) \right)  \right]\Psi(\br) \\
 & + \int \D{^2 \br} \left( \dou_i \Psi^\dagger(\br) \bV_i(\br) \Psi(\br) + \Psi^\dagger(\br) \bV^\dagger_i(\br) \dou_i \Psi(\br)\right) \Psi(\br)
\end{split}
\eeq
where 
\beq \label{SM:eqn:WVxVy}
\begin{split}
\bW(\br) = & -t\begin{pmatrix}
0 & 0 & \Delta z_{31}(\br) &   \Delta z_{41}(\br)  \\  
0 & 0 & \Delta z_{23}(\br) & \Delta z_{24}(\br) \\
\Delta z_{31}(\br) & \Delta z_{23}(\br)  & 0 & 0 \\
\Delta z_{41} (\br) & \Delta z_{24}(\br) & 0 & 0
\end{pmatrix} \\
\bV_1(\br) = & -t \begin{pmatrix}
0 & 0 & 0 & \Delta z_{1'4}(\br) \\
0 & 0 & 0 & 0  \\
0 & \Delta z_{3'2}(\br) & 0 & 0 \\
0 & 0 & 0 & 0
\end{pmatrix} \\
\bV_2(\br) = & -t \begin{pmatrix}
0 & 0 &  \Delta z_{1''3}(\br) & 0 \\
0 & 0 & 0 & 0  \\
0 &  0 & 0 & 0 \\
0 &  \Delta z_{4''2}(\br) & 0 & 0
\end{pmatrix} \end{split}
\eeq

\begin{figure}
    \centering
    \includegraphics[width=0.5\linewidth]{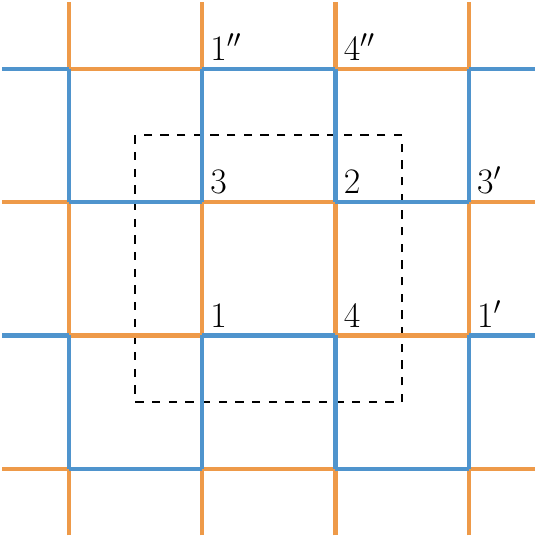}
    \caption{{\bf Derivation of Field Theory:} Schematic for the derivation of long wavelength disorder Hamiltonian.}
    \label{SM:fig:field_theory_disorder}
\end{figure}
(note $t=1$ is assumed below). The sites such as $1'$ and $1''$ correspond to those in the next neighbor unit cells, respectively, along 1 ($x$) and 2 ($y$) directions (see \figref{SM:fig:field_theory_disorder}). $\Delta z$ quantities are the random gauge fields induced by the flux disorder as described in \eqn{eqn:DiracF} of the main text. 
To proceed with the analysis, we will first reorganize the disorder term
\beq
\begin{split}
\mathcal{H}_F = & \int \D{^2 \br} \Psi^\dagger(\br) \bU(\br) \Psi(\br) \\
 & + \int \D{^2 \br} \left( \dou_i \Psi^\dagger(\br) \bV_i(\br) \Psi(\br) + \Psi^\dagger(\br) \bV^\dagger_i(\br) \dou_i \Psi(\br)\right) \Psi(\br)
 \end{split}
\eeq
where
\beq\label{SM:eqn:LocalDis}
\bU(\br) := \bW(\br) + \sum_i \left( \bV_i(\br) + \bV^\dagger_i(\br)  \right)
\eeq
We further define
\beq
\begin{split}
\bU(\br) &= \overline{\bU} + \bu(\br) \\
\bV_i(\br) & = \overline{\bV}_i + \bv(\br)
\end{split}
\eeq
where the $\overline{(~)}$ quantities are the disorder averages and the small letter quantities are fluctuations about the average.
Thus, we will take the starting point of the analysis as 
\beq\label{SM:eqn:HFreorg}
\begin{split}
{\cal H}_F = & \int \D{^2 \br} \Psi^\dagger(\br) \overline{\bU} \Psi(\br) \\
 & + \int \D{^2 \br} \left( \dou_i \Psi^\dagger(\br) \overline{\bV}_i \Psi(\br) + \Psi^\dagger(\br) \overline{\bV}^\dagger_i(\br) \dou_i \Psi(\br)\right) \Psi(\br) \\
 & + \int \D{^2 \br} \Psi^\dagger(\br) \bu(\br)  \Psi(\br)
\end{split}
\eeq
It can be shown that for any type of defect cluster distribution, terms with the averages will contribute to the change of the Dirac velocity by a quantity proportional to $\conc$. The main physically important quantity is the randomly fluctuating disorder is captured by the $4\times4$ matrix field $\bu(\br)$.

It is convenient to choose a basis in which 
\beq
\alpha_i = \sigma_1 \otimes \sigma_i.
\eeq
This is achieved by a unitary transformation described by the matrix
$$
\left(
\begin{array}{cccc}
 \frac{1}{2}-\frac{i}{2} & -\frac{1}{2}-\frac{i}{2} & 0 & 0 \\
 -\frac{1}{2}-\frac{i}{2} & \frac{1}{2}-\frac{i}{2} & 0 & 0 \\
 0 & 0 & \frac{1}{2}+\frac{i}{2} & \frac{1}{2}-\frac{i}{2} \\
 0 & 0 & \frac{1}{2}-\frac{i}{2} & \frac{1}{2}+\frac{i}{2} \\
\end{array}
\right)
$$
and the resulting field theory can be written as
\beq \label{SM:eqn:HField}
{\cal H}_0 + {\cal H}_F = \int \D{^2 \br} \begin{pmatrix}
0 & {\cal D} \\
{\cal D}^\dagger & 0
\end{pmatrix} \Psi^\dagger(\br)  \Psi(\br)
\eeq
with
\beq \label{SM:eqn:HField_detailed}
 {\cal D} = (1- f \conc) \sigma_i (-\ci \dou_i) + \sigma_i A_i(\br) + \tau_i m_i(\br) 
\eeq
where $i$ runs over $1,2$, we have defined
\beq
\tau_1 = \sigma_0, \tau_2 = \sigma_3,
\eeq
and $f$ is a number of order unity that depends on the type of defect cluster.
The random gauge fields $A_i(\br)$ and the mass terms $m_i(\br)$ are determined via 
\beq
\begin{split}
A_{1}(\br) &= u_{14}(\br) + u_{23}(\br) \\
A_{2}(\br) &= u_{13}(\br) - u_{24}(\br) \\
m_1(\br) & = u_{13}(\br) +  u_{24}(\br) \\
m_2(\br) & = -u_{14}(\br) + u_{23}(\br)
\end{split}
\eeq
where $u_{ab}$ are the matrix elements of $\bu(\br)$ introduced in \eqnref{SM:eqn:HFreorg}.

In the previous works~\cite{HatsugaiWen1997,Guruswamy2000}, the quantities $\bu(\br)$ considered are such that they are spatially uncorrelated random fields, and further that the $A_i$ are uncorrelated with $m_i$. We emphasize that this is not adequate to describe the flux disorder, and in this case, all the disorder fields are correlated with each other.

\section{RG Calculations}\label{SM:sec:RG}
For zero energy states we can treat $\mathcal{D}$ (see \eqref{SM:eqn:HField_detailed}) as a Hamiltonian with imaginary gauge fields, real mass and imaginary potential. Treating $\mathcal{D}$ as Hamiltonian and ignoring the change in the Dirac velocity, we write the action for the system as 
\begin{equation}
    \begin{split}
        S = & \int \frac{\D{^2\br}}{2\pi} \left[\Psi^\dagger(\br)\sigma_i(-\ci \partial_{i} + A_{i}(\br))\psi(\br) \right.\\
        & \left.+ m_1(\br)\Psi^\dagger(\br)\sigma_0\Psi(\br) + m_2(\br)\Psi^\dagger(\br)\sigma_3\Psi(\br)  \right].
    \end{split}
\end{equation}
Here $\Psi(\br)$ is a two component field unlike four component field we had in Eq. \ref{SM:eqn:HField} . To make the action more suitable for renormalization group analysis, the coordinate is changed from $(x,y)$ to $(z,\bar{z})$ where $z = x + iy$ and $\bar{z} = x - iy$. We further transforms the fields as follows, 

\begin{equation}    
    \Psi = \begin{pmatrix}
        \psi_1 \\ 
        \psi_2
    \end{pmatrix} \rightarrow 
    \begin{pmatrix}
        \psi \\ 
        \bar{\psi}
    \end{pmatrix}
\end{equation}

\begin{equation}
\begin{split}
     A_z = A_1 - \ci A_2, \quad A_{\bar{z}} = A_1 + \ci A_2 \\ 
     m_z = m_1 + \ci m_2, \quad m_{\bar{z}} = m_1 - \ci m_2.
\end{split}
\end{equation}
After the transformation, the action becomes
\begin{equation}\label{SM:eqn:raw_action}
    \begin{split}
        S = & \int \frac{\D{^2\br}}{2\pi} \left[\psi^\dagger(z)(\partial_{\bar{z}} + A_{\bar{z}})\psi(z) + \bar{\psi}^\dagger(\bar{z})(\partial_{z} + A_{z}(\br))\bar{\psi}(\bar{z})\right. \\ 
        & \left. + m_{z}(\br)\bar{\psi}^\dagger(\bar{z})\psi(z) + m_{\bar{z}}(\br)\psi^\dagger(z)\bar{\psi}(\bar{z})  \right]
    \end{split}
\end{equation}
In the above action, the fields $A_{\bar{z}}, A_{z}, m_{z}$ and $m_{\bar{z}}$  are gaussian distributed random variables. For brevity, only three probability density functions are shown below
\begin{equation}\label{SM:eqn:PD_disord}
    \begin{split}
    P[m_{z}, m_{\bar{z}}] &\sim \exp\left[ -\frac{1}{g_{m\bar{m}}} \int \frac{d^2\br}{2\pi}m_{\bar{z}}(\br)m_{z}(\br) \right] \\ 
    P[A_z, A_{\bar{z}}] &\sim \exp\left[ -\frac{1}{g_{j\bar{j}}} \int \frac{d^2\br}{2\pi}A_{\bar{z}}(\br)A_{z}(\br) \right] \\
    P[A_z, m_{\bar{z}}] &\sim \exp\left[ -\frac{1}{g_{j\bar{m}}} \int \frac{d^2\br}{2\pi}m_{\bar{z}}(\br)A_{z}(\br) \right].
    \end{split}
\end{equation}
The last term in the \eqnref{SM:eqn:PD_disord} accounts for the correlation between mass and gauge fields that are present in our system. Any physical quantity calculated from the action in \eqnref{SM:eqn:raw_action} needs to be disorder averaged. Instead of performing averaging in the physical quantity, using supersymmetry technique, we obtain a disorder averaged action~\cite{Guruswamy2000,Efetov1996} from which disorder averaged quantities can be calculated directly. In the supersymmetry technique, we facilitate disorder averaging by adding a bosonic field with exact same form of the original action but with bosonic fields $\phi$ and $\bar{\phi}$.
The disorder averaged action we obtained is as follows 
\begin{equation}
        \begin{split}
        S_{\text{eff}} = & \int \frac{\D{^2\br}}{2\pi} \left[\psi^\dagger(z)\partial_{\bar{z}}\psi(z) + \bar{\psi}^\dagger(\bar{z})\partial_{z}\bar{\psi}(\bar{z}) \right.  \\ 
        & + \phi^\dagger(z)\partial_{\bar{z}}\phi(z) + \bar{\phi}^\dagger(\bar{z})\partial_{z}\bar{\phi}(\bar{z}) \\
        &+ g_{jj}\mathcal{O}_{jj} + g_{j\bar{j}}\mathcal{O}_{j\bar{j}} + g_{\bar{j}\bar{j}}\mathcal{O}_{\bar{j}\bar{j}}\\ 
        & \left.+ g_{mm}\mathcal{O}_{mm} + g_{m\bar{m}}\mathcal{O}_{m\bar{m}} + g_{\bar{m}\bar{m}}\mathcal{O}_{\bar{m}\bar{m}}  + ... \right]
        \end{split}
\end{equation}
where $\mathcal{O}_{\alpha\beta} = \mathcal{O}_{\alpha}\mathcal{O}_{\beta}$ and $g_{\alpha\beta}$ are the correlation between the coefficients of operators $\mathcal{O}_\alpha$ and $\mathcal{O}_\beta$. The operators $\mathcal{O}_\alpha$'s are listed below,
\begin{equation}
    \begin{split}
    \mathcal{O}_j &= \left( \psi^\dagger(z)\psi(z) + \phi^\dagger(z)\phi(z) \right) \\
    \mathcal{O}_m  &= \left( \bar{\psi}^\dagger(\bar{z})\psi(z) + \bar{\phi}^\dagger(\bar{z})\phi(z) \right)
    \end{split}
\end{equation}
and $\mathcal{O}_{\bar{j}}$ and $\mathcal{O}_{\bar{m}}$ are conjugate of $\mathcal{O}_j$ and $\mathcal{O}_m$ respectively. The above action describe a system of interacting fermions and bosons with interaction strength given by the coupling parameters of the disorder fields in the original system (Eqn. \ref{SM:eqn:HField_detailed}). Using the operator product expansions of the operators $\mathcal{O}_{\alpha\beta}$, we obtain the RG flow equations for the coupling parameters $g_{\alpha\beta}$ upto one loop order. There are in total ten independent $g_{\alpha\beta}$ parameters and we have obtained the RG flow equations for all of them. For brevity, only three of them are shown below 
\begin{equation}\label{SM:eqn:gjj}
    \frac{\D g_{j\bar{j}}}{\D l} = g_{m\bar{m}}^2 - 2|g_{j\bar{m}}|^2
\end{equation}
\begin{equation}\label{SM:eqn:gmm}
    \frac{\D g_{m\bar{m}}}{\D l} = - |g_{j\bar{m}}|^2
\end{equation}
\begin{equation}\label{SM:eqn:gjm}
    \frac{\D g_{j\bar{m}}}{\D l} =  - g_{j\bar{j}}g_{j\bar{m}}.
\end{equation}
Equations \ref{SM:eqn:gjj} and \ref{SM:eqn:gmm} have been extensively studied in the literature~\cite{HatsugaiWen1997, Guruswamy2000}. Based on these two equations it has been claimed that in presence of mass disorder, freezing multifractality of the zero energy state is inevitable~\cite{HatsugaiWen1997}. However, when the correlation $g_{j\bar{m}}$ is added, the RG flow is affected significantly. In fact for the special case, $g_{j\bar{j}} = g_{m\bar{m}}=g_{j\bar{m}}$, the parameters flows to zero, making the disorders irrelevant. For other generic cases, the RG flow initially may suppress the strength of the coupling parameters (see \figref{fig:RGresults}), but eventually the $g_{m\bar{m}}$ parameter saturates to a finite value and monotonically increases the $g_{j\bar{j}}$ parameter under RG flow, taking the system to strong disorder limit. But the length scale at which freezing takes place can be significantly increased in the presence of the $g_{j\bar{m}}$.
\section{$T$-matrix Calculations}\label{SM:sec:Tmatrix}
The disorder in our random $\pi$ flux model is non-perturbative, making the RG calculation unreliable. To circumvent the issue and with motivation to go beyond the Dirac regime, we calculated the spectral function using the T-matrix method. The crux of the T-matrix method is to consider the scattering processes where scattering takes place at the same impurity. This assumption allows us to obtain disorder-averaged Green's function in the presence of an impurity at low concentration~\cite{Ostrovsky2006}. To proceed, we start with Green's function of the clean system, 
\begin{equation}
    G(\bk;z) = \left[z - H_0(\bk)  \right]^{-1}
\end{equation}
where, $H_0(\bk)$ is the $4\times4$ $\bk$-space hamiltonian of the clean system. If we add disorder into this clean system the real space hamitonian will change as 
\begin{equation}
    H = H_0 + V
\end{equation}
where $V$ is same as $H_F$ in \eqref{eqn:DiracF}. $V$ can be expressed in terms of change in the hopping amplitude. To make the analysis simpler, we have considered defects which can be introduced by changing the sign of the hopping amplitudes only inside the unitcells. Restricting ourselves to this particular types of disorder, we express $V$ as
\begin{equation}
    V = \sum_{\alpha} \ket{\br_\alpha}\Gamma_{\alpha\alpha}\bra{\br_{\alpha}} 
\end{equation}
where $\alpha$ labels the defects. $\ket{\br_{\alpha}}$ is a four component vector $(\ket{\br_{\alpha}^1}, \ket{\br_{\alpha}^2}, \ket{\br_\alpha^3}, \ket{\br_\alpha^4})^T$ where $\ket{\br_\alpha^i}$ denotes the state corresponding to the $i$-th site of the unit cell located at $\br_\alpha$. $\Gamma_{\alpha\alpha}$ are  $4\times4$ matrices like $\bW(\br)$ in  \eqnref{SM:eqn:WVxVy}. In the $\bk$ space,
\begin{equation}
    V = \frac{1}{N}\sum_{\alpha}\sum_{\bk_1\bk_2}\left[e^{i(\bk_1 - \bk_2).\br_\alpha}\right] \ket{\bk_1}\Gamma_{\alpha\alpha}\bra{\bk_2}
\end{equation}
where $\ket{\bk}$ is again a four component vector defined like $\ket{\br_\alpha}$ vector. Anticipating the importance of multi defect scattering, we club $n$ number of defects together and treat them as \emph{a single impurity}. We are not assuming that the defects in an impurity need be close to each other. Keeping this in mind, we decompose the $\br_\alpha$ as 
\begin{equation}
    \br_\alpha = \br + \bl_\alpha
\end{equation}
and $V_{\bk_1\bk_2}$ is rewritten in terms of $\br$ and $\bl_\alpha$
\begin{equation}
\begin{split}
    V_{\bk_1\bk_2} &= \frac{1}{N}\sum_{\alpha}\left[e^{i(\bk_1 - \bk_2).(\br + \bl_\alpha)}\Gamma_{\alpha\alpha}\right] \\
    & = \sum_{\alpha}W_{\bk_1}^{*\alpha}\Gamma_{\alpha\alpha}W_{\bk_2}^{\alpha}e^{i(\bk_1 - \bk_2)\cdot\br}
\end{split}
\end{equation}
where, 
\begin{equation}
    W_{\bk}^{\alpha} = \frac{1}{\sqrt{N}}e^{-i\bk.\bl_\alpha}.
\end{equation}
The Green's function of the disordered system can be expressed in terms of the Green's function of the clean system $G$, and the $V$ using Dyson equation 
\begin{equation}
    \mathcal{G} = G + GVG + GVGVG+ \ldots
\end{equation}
This equation can be written in compact form 
\begin{equation}
    \mathcal{G} = G + GTG
\end{equation}
where, 
\begin{equation} \label{SM:eqn:Tmatdef}
    T = V + VGV + VGVGV + \ldots
\end{equation}
$T$ is called the $T$ matrix corresponding to the perturbation $V$. Calculating the $T$ matrix considering all the impurities is an extremely difficult problem. However, $T$ matrix method relies on the assumption that the scattering to a single defect is enough. Using the definition in \eqnref{SM:eqn:Tmatdef}, we write the $T$ matrix for a single impurity at $\br$ is 
\begin{equation}
\begin{split}
    T_{\bk\bk'}(\br) =&  V_{\bk\bk'} + V_{\bk\bk_1}G(\bk_1;z)V_{\bk_1\bk'}  \\
    &+ V_{\bk\bk_1}G(\bk_1;z)V_{\bk_1\bk_2}G(\bk_2;z)V_{\bk_2\bk'} + \ldots
\end{split}
\end{equation}
Writing $V_{\bk\bk'}$ explicitly we get
\begin{equation}
\begin{split}
    T_{\bk\bk'}(\br) &= e^{i(\bk - \bk').\br}\left[W_{\bk}^{*\alpha}\Gamma_{\alpha\alpha} W_{\bk'}^{\beta}\delta_{\alpha,\beta} \right. \\&+ \sum_{\bk_1} W_{\bk}^{*\alpha} \Gamma_{\alpha\alpha} W_{\bk_1}^{\alpha} G(\bk_1;z) W_{\bk_1}^{*\beta} \Gamma_{\beta\beta} W_{\bk'}^\beta \\
    &+\sum_{\bk_1\bk_2}W_{\bk}^{*\alpha}\Gamma_{\alpha\alpha} W_{\bk_1}^{\alpha} G(\bk_1;z) W_{\bk_1}^{*\alpha_1}\Gamma_{\alpha_1\alpha_1} \\ 
    &\left.W_{\bk_2}^{\alpha_1} G(\bk_2;z) W_{\bk_2}^{*\beta}\Gamma_{\beta\beta} W_{\bk'}^\beta  + \ldots \right]\\ 
\end{split}
\end{equation}
The position $\br$ of the impurity is a random variable with uniform probability distribution. So, the disorder averaged $T$ matrix can be obtained as follows
\begin{equation}
    \disave{T_{\bk\bk'}} = \sum_{\br} \frac{1}{N}T_{\bk\bk'}(\br)
\end{equation}
where $N$ is the total number of unit cells in the system. Performing the disorder averaging we obtain
\begin{equation}
\begin{split}
    \disave{T_{\bk\bk'}} =& W_{\bk}^{*\alpha}\Gamma_{\alpha\alpha} W_{\bk'}^{\beta}\delta_{\alpha,\beta}\delta_{\bk,\bk'} \\
    &+ \sum_{\bk_1} W_{\bk}^{*\alpha} \Gamma_{\alpha\alpha} W_{\bk_1}^{\alpha} G(\bk_1;z) W_{\bk_1}^{*\beta} \Gamma_{\beta\beta} W_{\bk'}^\beta \delta_{\bk,\bk'} \\
    &+\sum_{\bk_1\bk_2}W_{\bk}^{*\alpha}\Gamma_{\alpha\alpha} W_{\bk_1}^{\alpha} G(\bk_1;z) W_{\bk_1}^{*\alpha_1}\Gamma_{\alpha_1\alpha_1} W_{\bk_2}^{\alpha_1}\\
    &G(\bk_2;z) W_{\bk_2}^{*\beta}\Gamma_{\beta\beta} W_{\bk'}^\beta \delta_{\bk,\bk'}  + \ldots \\ 
\end{split}
\end{equation}
Notice that $\disave{T_{\bk\bk'}}$ is diagonal in $\bk$ showing that translation invariance is restored upon disorder averaging. From here onward, we will write $\disave{T_{\bk}}$ instead of $\disave{T_{\bk\bk'}}$. Identifying the geometric progression we rewrite the term as
\begin{equation}
\begin{split}
    \disave{T_{\bk}}&= W_{\bk}^{*\alpha} \delta_{\alpha,\beta}\Gamma_{\beta\beta} W_{\bk}^{\beta} + W_{k}^{*\alpha}\left[ \sum_{m=1} (\Gamma\Pi)^m\right]_{\alpha\beta}\Gamma_{\beta\beta}W_{k}^\beta \delta_{\bk,\bk} \\ 
    & = W_{k}^{*\alpha}\left[ \sum_{m=0} (\Gamma\Pi)^m\right]_{\alpha\beta}\Gamma_{\beta\beta}W_{k}^\beta  \\ 
    & = W_{k}^{*\alpha}\left[\mathbb{I}  - \Gamma\Pi\right]^{-1}_{\alpha\beta}\Gamma_{\beta\beta}W_{k}^\beta \\ 
    & = W_{k}^{*\alpha} \mathcal{F}_{\alpha\beta} W_{k}^\beta \delta_{\bk,\bk}
\end{split}
\end{equation}
where $\Pi$ is defined as
\begin{equation}
    \Pi_{\alpha\beta} = \sum_{\bk}W_{\bk}^{\alpha}G(\bk)W_{\bk}^{*\beta}
\end{equation}
and $\mathcal{F}$ as
\begin{equation}
    \mathcal{F}_{\alpha\beta} = \left[\mathbbm{I}  - \Gamma\Pi\right]^{-1}_{\alpha\beta}\Gamma_{\beta\beta}.
\end{equation}
Once we have the $T$ matrix for a single impurity, we can get the disorder averaged Green's function simply as 
\begin{equation}
    \mathcal{G}(\bk,\bk';z) = G(\bk;z)\delta_{\bk\bk'} + G(\bk;z)\disave{T_{\bk\bk'}}G(\bk';z).
\end{equation}
However, in a real system, there are multiple impurities and if the concentration $c$ of impurities are small enough we can approximately write the disorder averaged self energy 
\begin{equation}
    \Sigma_{\bk\bk'} = c\disave{T_{\bk}}.
\end{equation}
In terms of self-energy, the disorder-averaged Green's function is
\begin{equation}\label{SM:eqn:selfenrg}
    \mathcal{G}(\bk;z)^{-1} = G(\bk;z)^{-1} + \Sigma_{\bk}.
\end{equation}
Using this Green's function, it's straightforward to calculate spectral function 
\begin{equation}\label{SM:eqn:specfunc}
    A(\bk;\omega) = -\frac{1}{\pi}\lim_{\eta\to0}\mathrm {Tr}\left[\text{Im}\left[(\mathcal{G}(\bk;\omega + i\eta)\right]\right]
\end{equation}
and density of states
\begin{equation}\label{SM:eqn:dos}
    \rho(\omega) = \frac{1}{N}\sum_{\bk}A(\bk;\omega).
\end{equation}

\begin{figure}
    \centering
    \centerline{\includegraphics[width=\linewidth]{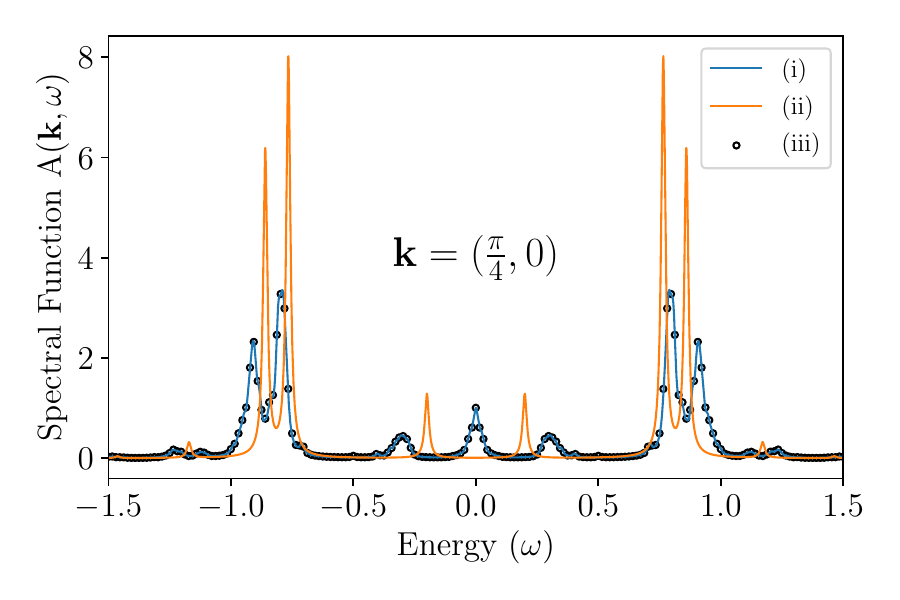}}
    \caption{{\bf Importance of coherent multi-defect scattering: } Comparison between the spectral functions ($A(\bk;\omega)$) corresponding to the two defect clusters scattering and single defect cluster scattering in a system of size $16\times16$. (i) Spectral function calculated numerically by putting two Greek cross clusters separated by $\br$ and averaging over all possible values of $\br$. (ii) $A(\bk;\omega)$ calculated using $T$ matrix method for a single Greek cross cluster. (iii) $A(\bk;\omega)$ calculated using $T$ matrix method for two 
    Greek cross defects averaged in the same way as (i).}
    \label{SM:fig:mult_scat_imps}
\end{figure}

We have calculated the spectral function for two cases using T matrix method: one where a single impurity contain a single Greek cross defect, and in the other case, a single impurity contain two Greek cross defects. In the later case, the result is averaged over all possible disorder realizations. We can clearly see in the Fig. \ref{SM:fig:mult_scat_imps} that the former case doesn't get contribution to the zero energy  from $\bk = (\frac{\pi}{4}, 0)$. However, the later case does get the contribution, evidenced from the peak at zero energy. This clearly demonstrate that scattering to a single defect is inadequate to give rise to zero energy states.
However, one might think whether consideration of two-defect scattering is sufficient to describe the zero energy state. The answer is no. From numerical computation we know that the Greek cross defect does produce a singularity in the density of state at zero energy. However, the DOS calculated using the \eqnref{SM:eqn:selfenrg} and \eqnref{SM:eqn:dos} doesn't have peak at the zero energy.
As evident from these calculations that the zero energy states is a result of collective scattering from a large (possibly thermodynamically large) number of defects and pose a serious challenge to the study of random $\pi$-flux model analytically.

\end{document}